\pdfoutput=1
\documentclass[fleqn,usenatbib]{mnras}

\usepackage{newtxtext,newtxmath}
\usepackage[T1]{fontenc}

\DeclareRobustCommand{\VAN}[3]{#2}
\let\VANthebibliography\thebibliography
\def\thebibliography{\DeclareRobustCommand{\VAN}[3]{##3}\VANthebibliography}

\usepackage{gensymb}    %Inclusion of degree symbol
\usepackage{graphicx}	% Including figure files
\usepackage{amsmath}	% Advanced maths commands
\usepackage{comment}    %comment out a section of the writing
\usepackage{rotating}   %rotate a table
\usepackage{hyperref}   %hyperlink
\usepackage{caption}
\usepackage{subcaption}
\usepackage{soul}
\usepackage[normalem]{ulem}
\usepackage{dirtytalk}
\usepackage[dvipsnames]{xcolor}
\usepackage{float}
\usepackage[dvipsnames]{xcolor}
\usepackage{verbatim}
\title[GARCIA II: Kinematic modelling and Mass modelling ]{The GMRT archive atomic gas survey - II.  Mass modelling and dark matter halo properties across late-type spirals }

\author[Biswas et al.]{
Prerana Biswas$^{1,2}$ \thanks{E-mail: prerana.biswas1994@gmail.com},
Veselina Kalinova$^{3}$,
Nirupam Roy$^{2}$,
Narendra Nath Patra$^{4}$,
Nadezda Tyulneva$^{3}$
\\
$^{1}$Joint Astronomy Programme, Indian Institute of Science, Bangalore 560012, India \\
$^{2}$Department of Physics, Indian Institute of Science, Bangalore 560012, India\\
$^{3}$Max Planck Institute for Radioastronomy, Auf dem Hügel 69, D-53121 Bonn, Germany \\
$^{4}$Department of Astronomy, Astrophysics and Space Engineering, Indian Institute of Technology Indore, Indore 453552, India\\
}
\date{Accepted XXX. Received YYY; in original form ZZZ}

\pubyear{2015}

\begin{document}
\label{firstpage}
\pagerange{\pageref{firstpage}--\pageref{lastpage}}
\maketitle

\begin{abstract}
 Studying the kinematics and mass modelling of galaxies from H~{\sc i} 21 cm data provides valuable insights into the properties of both the baryonic components and the dark matter halo in nearby galaxies.  Despite many observational studies,  mass modelling of galaxies remains challenging due to different limitations. For example, most of the previous studies involving mass modelling are based on rotation curves derived from two-dimensional velocity fields from H~{\sc i} or H$\alpha$ spectroscopic observation which are often affected by beam smearing and projection effect. However,  kinematic modelling done by fitting the "Tilted ring model" to three-dimensional data cube is not affected by these issues. In this study, we present and compare 3D kinematic modelling of a pilot sample of eleven galaxies from the GMRT archive atomic gas survey (GARCIA) using two different publicly available pipelines.  We model the observed H~{\sc i} rotation curve using 3.6 $\mu$m infrared data and SDSS r-band data for stellar contribution,  H~{\sc i} surface density profile for gas, and Navarro-Frenk-White (NFW) profile for dark matter halo; and employ the Markov Chain Monte Carlo (MCMC) optimization method for parameter estimation. Further, to validate our analysis, we revisit important scaling relations, e.g., the M$_{gas}$-M$_{star}$ relation, M$_{star}$-M$_{halo}$ relation, M$_{gas}$-M$_{halo}$ relation and Baryonic Tully-Fisher relation (BTFR). The scaling relations from our analysis are broadly consistent with that reported in the literature. A larger sample of galaxies from GARCIA in the near future will allow studying these scaling relations in greater details.

%Our results align with these relations, except for BTFR, which shows a hint of a shallower slope. This can not be affirmed with our sample of 11 galaxies; the larger sample from GARCIA will be required to state if it is statistically significant.}} 
\end{abstract}

\begin{keywords}
galaxies: general - galaxies: kinematics and dynamics - galaxies: stellar content - galaxies: haloes - galaxies: fundamental parameters
\end{keywords}

%%%%%%%%%%%%%%%%%%%%%%%%%%%%%%%%%%%%%%%%%%%%%%%%%%

%%%%%%%%%%%%%%%%% BODY OF PAPER %%%%%%%%%%%%%%%%%%

\section{Introduction}
\label{intro}
  Since the last four decades, it has been well observed that the rotation curves of galaxies suggest a large discrepancy between dynamically determined mass and mass measured from the luminous matter \citep{bosma_1978, rubin1978, rotc_bosma1981b, rotcur1, sofue_2001}. This astrophysical evidence invoked the idea of introducing dark matter in galaxies. Although there are other studies to explain the missing mass, e.g., Modified Newtonian Dynamics (MOND) formalism \citep{mond_milgrom1983, mass_mod1_begeman1991, mass_mod2_sanders1996}, Scalar tensor vector gravity theory  (STVG) \citep{stvg_theory}, Modified Gravity (MOG) theory \citep{mog_theory}, non-local gravity theory \citep{nonlocal_gravity1, mond_mog_nonlocal}, to date, the most accepted one is through $\Lambda$CDM model, i.e., the introduction of a dark matter halo in the galaxies. Different models for dark matter distribution inside the galaxies have been proposed in the literature. These are either motivated by observation or simulation. Among these, the most popular and successful ones include the following: observationally motivated pseudo-isothermal (pISO) halo \citep{mass_mod1_begeman1991}, which assumes that the halo is spherical with a constant core density and density profile approximately equal to an isothermal sphere; Burkert profile \citep{Burkert_halo_1995} which is modified version of pISO profile and diverges slowly at large radii in comparison to pISO profile; Navarro-Frenk-White (NFW) profile \citep{navarro_1996} inferred from the $N$-body dark matter only simulations of structure formation; DC14 profile \citep{DC14_halo_profile} inferred from cosmological hydrodynamic simulation of galaxy formation and considered the effect on halo due to accretion of baryonic matter within the halo; etc.
  
  Our knowledge of the mass distribution of different components in galaxies is mostly based on the study of the rotation curve. Large samples of nearby galaxies have had their kinematics measured over the past few decades, either in  optical with Fabry-Perot interferometry of the H$\alpha$ line \citep[ e.g., the GHASP sample of $\sim$ 200 galaxies,][]{ghasp_rotc_epinat2008} or using interferometric data of H~{\sc i} 21-cm line emitted during the hyperfine transition of atomic hydrogen  \citep{rotc_bosma1981b, mass_mod1_begeman1991, broeils1992, mass_mod2_sanders1996, whisp_main2001, verheijen2001, things_main, mcgaugh2012}. However, to extract the rotation curve, it is beneficial to use the H~{\sc i} interferometric data, as the H~{\sc i} disk extends far larger than the stellar disk, and we can probe the velocity up to a larger radius. Also, sometimes the rotation curves from other wavelengths do not reach the flat part and thus can not probe the region where the dark matter usually dominates the dynamics. Besides that, since H~{\sc i} gas is cold in comparison to other components and follows nearly circular orbits (with typical dispersion velocity $\sim$ 10 kms$^{-1}$ for normal galaxies \citep{things_main} and $\sim$ 20 kms$^{-1}$ for dwarfs \citep{little_things_mass_model}), it traces the gravitational potential more precisely.   To date, for most of the studies, the 2D method, i.e., fitting the 2D Tilted-ring model to the 2D velocity field, is used to extract the kinematics, but this method suffers the problem of beam smearing and projection effect. \citet{bosma_1978} showed that when the data have less than seven beams inside the Holmberg radius, the slope of the inner rotation curve is biased towards low velocity due to the beam smearing effect. Further, \citet{sancisi_1979} and \citet{sofue_2001} showed that due to the severe projection effect, the reliable rotation curve could not be obtained for galaxies with very high inclination. Besides that, \citet{fat} showed that there are clear systemic differences between 2D fitting and 3D fitting, i.e., fitting the 3D Tilted ring model to 3D data cube using their own developed Tilted ring fitting pipeline, FAT (Fully Automated TiRiFiC, Tilted Ring Fitting Code). Their study suggests the importance of doing 3D kinematic modelling over 2D modelling. With the 3D kinematic modelling over the observed data cube,  these effects can be neglected, and an accurate rotation curve can be obtained.
  
  Despite a large amount of kinematic observations and different halo models, there are still challenges in modelling the mass distribution of different components in galaxies. The large and different scatters in the mass-to-light ratio in different bands make it difficult to estimate the distribution of the stellar mass accurately. Besides that, the 2D method in obtaining rotation curves previously limited these studies to galaxies with very high or very low inclination. 
  
  In the current paper, we derive the 3D kinematic modelling and mass modelling of the eleven galaxies from the GMRT ARChIve Atomic gas survey (GARCIA) \citep{biswas2022}. The GARCIA survey aims to explore the potential of the GMRT archive by building a sample of 515 nearby galaxies for which the H~{\sc i} interferometric spectral line data is available with an adequate signal-to-noise ratio. The objective of the survey is to analyze this entire data set uniformly in a homogeneous manner and explore various science cases. As a part of this project, our first paper in this series, GARCIA-I, presents the data products of a pilot sample, and here in this second paper of the series, the kinematics and mass modelling are presented.
  
  We derive the 3D kinematic modelling with the available pipelines FAT \citep{fat} and BBarolo \citep{BBarolo} of fitting 3D Tilted ring models \citep{tiltd_ring1974} and compare the results found from both the procedure. In most of the previous work involving mass modelling, the H~{\sc i} interferometric data that have been used are gathered from different observations and the data analysis is done following different procedures by different groups, creating inhomogeneity in data products. However, here we use the homogeneous data products from the GARCIA survey \citep{biswas2022}. We further compare the properties of these sources found from the 3D kinematic modelling with those from single-dish and optical observations.
  
  As the large and different scatter of mass-to-light ratio in different bands makes the mass modelling of the galaxies most challenging to accurately determine the stellar mass and hence the best-fitted model,  we use optical and infrared photometric data to derive the luminosity distribution using Multi-Gaussian Expansion (MGE) procedure \citep{mge2, mge3, mge_cappellari2002}. These photometric profiles are further used to find out the velocity component of the stellar disk through Jeans Anisotropic Model (JAM) \citep{jam_cappellari_2008} assuming a constant mass-to-light ratio (M/L) with radius. To determine the contribution of gas to the total velocity, we use the surface density from the 3D kinematic modelling. As mentioned above, most of the studies of mass modelling are based on the rotation curve derived from the 2D methods. There are only a handful of studies where the mass modelling is done using the rotation curve derived in 3D kinematic modelling  \citep{susham_superthingal_massmod, sushma_mass_model, Shelest2020_3drot_masmod, Teodoro2022_3Dmass_model_mcmc}. \citet{sushma_mass_model} showed that the rotation curve obtained in the 3D method gets best fitted when we use the NFW profile to model the halo, at least for dwarf galaxies. Thus, in this paper, we use the NFW profile for modelling the halo \citep{navarro_1996,navarro_1997}. We use the Markov Chain Monte Carlo (MCMC) optimization method \citep[e.g.,][]{sivia1998} to find out the best-fitted model of the rotation curve of the galaxies.

    To check the consistency of the derived quantities from the kinematic and mass modelling, we revisit some of the most fundamental scaling relations for galaxies, i.e., the M$_{gas}$-M$_{star}$ relation \citep{Parkash2018}, M$_{star}$-M$_{halo}$ relation \citep{moster2013}, M$_{gas}$-M$_{halo}$ \citep{Padmanabhan2017} relation and the Baryonic Tully-Fisher relation (BTFR) \citep{Tully_1977, McGaugh2000}. Among these relations, we emphasised our interest in the BTFR as in most of the previous studies of BTFR, 2D kinematic modelled velocities are used \citep{BFTR1}, while we used velocities from 3D kinematic modelled data. The BTFR is a vital relation correlating the baryonic mass and the rotation velocity of the galaxies. It has been extensively used to estimate the distances \citep[e.g.,][]{Tully_1977}, determine the value of the Hubble constant \citep[e.g.,][]{h0_tf1, h0_tf2}, study the local galaxy flows \citep[e.g.,][]{local_flow_ft}, examine the galaxy formation models in a $\Lambda $CDN cosmology \citep[e.g.,][]{gal_for_tf1, gal_for_tf2}, and test theories like modified gravity \citep[e.g.,][]{mond_milgrom1983, sanders_1990, mcgaugh2012}. So far, different measures of rotation velocity have been used to establish this relation. That includes  $w_{p20}$ and  $w_{m50}$, the width of the global H~{\sc i} profile obtained at the $20\%$ of the peak and at the $50\%$ of the mean flux, respectively; $V_{flat}$, the average circular velocity along the flat part of the rotation curve; $V_{max}$,  peak velocity in the rotation curve etc. As $w_{p20}$ and  $w_{m50}$ are relatively easier to obtain from the single-dish observation, in most of the previous studies of this relation, the widths of H~{\sc i} spectra are used as the tracer of the rotation velocity. However, as the global H~{\sc i} spectra are the specially integrated flux,  the widths of it may not be an accurate measurement for rotation velocity.  If a galaxy has a close neighbour or an interacting one, then the large beam size of the single-dish may not resolve the system and will result in a larger and inaccurate spectral width.  \citet{BFTR1} showed that $V_{flat}$ obtained from the kinetically modelling gives the tightest BTFR. Besides that, in most of the previous cases, the inclinations used for correcting the widths of the spectra are derived from the optical disk \citep[e.g.,][]{McGaugh2000} or from 2D kinematic modelling \citep{BFTR1}. The inclination angles of the galaxies, which is the dominant error \citep{dominant_err_in_BTFR, dominant_err_in_BTFR2} in the BTFR, may sometimes differ significantly between what is defined from the optical disk or H~{\sc i} disk or found by doing the 2D or 3D  kinematic modelling of the H~{\sc i} interferometric data; these can lead to different scatter and possibly different slope and intercept in  BTFR.  Here, in this paper, we compare the optical inclinations with the kinematic inclination and inclination defined from moment zero maps (i.e., the integrated intensity over the velocity axis produced from 3D data cube). We also compare the velocity found from the H~{\sc i} spectra from single-dish and interferometric observations with the velocities found from 3D kinematic modelling. With these different measures of velocities from kinematic modelling and corrected widths for optical and kinematic inclination, we revisit the BTFR. 
    
    In section \ref{data}, we describe the selected sample and the data used for the kinematic and mass modelling; section \ref{sec:kinematics} demonstrates the results of the 3D fitting done through FAT and BBarolo; the comparison of inclination angles and rotation velocities found from different methods are described in section \ref{sec:comp_inc_vel}; section \ref{massmodel} contains the details of the procedure for doing mass modelling and discussion on the corresponding results. The notes on mass modelling for individual galaxies are  given in section \ref{sec:gal_note}.    In section \ref{sec:discussion}, we explore some of the important scaling relations, e.g., the M$_{star}$-M$_{halo}$ relation, the BTFR relation etc.   And finally, in section \ref{sec:conclusion}, we summarise our results and discuss the future aspects of this work. 

\section{Sample Selection and Data}
\label{data}

\subsection{H~{\sc i} interferometric data}
As mentioned in the introduction, we use the uniformly analyzed H~{\sc i} interferometric data of the pilot sample of eleven galaxies from the GARCIA Survey \citep{biswas2022}. Sample galaxies are bright ($m_B < 14.10$) nearby (distance $\le 46.96$ Mpc) spiral or irregular galaxies, each of which has a different combination of inclination and position angles. They exhibit a broad range of H~{\sc i} masses ($\sim 10^8M_{\sun}$ to $\sim 10^{10}M_{\sun}$) and  H~{\sc i} spectral widths ($\sim 55 kms^{-1}$ to $\sim 150 kms^{-1}$ ) . The sources are also well distributed in the parameter space of H~{\sc i} diameter versus H~{\sc i} mass \citep[see][figure 7]{biswas2022}.  Tables 1 and 2 from \citet{biswas2022} respectively represent the cross-identification names and the general properties of these sources.  The details of the data reduction are described in detail in \citet{biswas2022}. The data cubes from these analyses are used for doing the 3D kinematic modelling and hence to extract the rotation curve of these galaxies. The distances used in this study are not derived from the Tully-Fisher relation, except for NGC7741, as this sample will be further used to check if the parameters found in the mass modelling and kinematic modelling are consistent when placed in the existing Baryonic Tully-Fisher relation.   For the source NGC7741, the method of Sosies \citep[see][]{sosies}, which uses TF relation indirectly to calculate the distance, is used; this distance is consistent with the distance of the group associated with this galaxy. The distances used in this study and the method for determining these distances for all the sources are mentioned in table 2 of \citep{biswas2022}.

\subsection{Optical and Infrared data}
As mentioned in the introduction, the mass-to-light ratio of the galaxies varies significantly depending on the observing band. However, \citet{ml3.6_mcgaugh2014}, \citet{ml_3.6_norris2016}, \citet{ml3.6_schombert2019} from stellar population synthesis model and \citet{ml3.6_eskew2012}, \citet{ml3.6_zhang} from colour-magnitude diagram showed that for all galaxies M/L $\sim$ 0.5 in 3.6-micron and it is the tightest value of M/L in comparison to other wavelengths. Thus, we tended to obtain the stellar disk's photometric profile primarily using 3.6-micron infrared data. For this, we used Spitzer IRAC-3.6 micron data from the Spitzer Survey of Stellar Structure in Galaxies (S4G)  project \citep{s4g1,s4g2,s4g3} and Super mosaic data products from Spitzer Heritage Archive \citep{spitzer_heritage_arch}. But for some sources of our sample, the 3.6 $\mu m$ data is not readily available; hence we also use optical photometric data of SDSS r-band \citep{sdss_iv_overview} for the available sources. This also provided us with the opportunity to study if there is any systematic difference in mass modelling between using infrared data and optical data.  We derived photometric profiles from each available band for each galaxy  and used them separately to compute the mass modelling. The sources' names and the available data used for the stellar component are given in the table  \ref{tab:stellar_data}.

\begin{table} 
\caption{Data for deriving the stellar component of the galaxies. }  
\begin{center}
\begin{tabular}{cc}
\hline
     Spitzer IRAC-3.6 micron & NGC0784, NGC3027, NGC3359, \\
     from S4G project & NGC4068, NGC4861, NGC7497, \\
      & NGC7741, NGC7800 \\
      & \\
     Spitzer IRAC-3.6 micron & NGC1156 \\
      from Spitzer Heritage Archive & \\
      & \\
     SDSS r-band &  NGC0784, NGC3359, NGC4068,  \\
      &  NGC4861, NGC7292, NGC7497, \\
      & NGC7610, NGC7741, NGC7800 \\
     
     \hline
\end{tabular}
\end{center}
\label{tab:stellar_data}
\end{table}

\section{Kinematic modelling}
\label{sec:kinematics}
The most important factor in comprehending the mass distribution of galaxies is their kinematic modelling. The rotation curve obtained in kinematic modelling provides information on the distribution of dark matter inside galaxies when paired with star and gas velocities. Additionally, it gives us a precise estimate of the total mass of the galaxies, which is necessary to construct other significant scaling relations, such as the Baryonic Tully-Fisher relation. Further, the study of angular momentum \citep{anguler_mont3, anguler_momnt_sushma, bryn_spect_anguler_montm}, which is essential to understand galaxy formation and evolution, requires a precise determination of the rotation curve.

The most common approach for extracting the kinematics is Tilted-ring modelling \citep{tiltd_ring1974}. In this model, the galactic disk is assumed to consist of concentric rings, each having an individual inclination, position angle, and rotation velocity. The 3D method must be applied to obtain an accurate rotation curve and kinematic modelling.

To construct the 3D kinematic modelling for these eleven galaxies, we used both the recently developed pipelines, 3DBarolo: a new 3D algorithm to derive rotation curves of galaxies (BBarolo) \citep{BBarolo} and Fully Automated TiRiFiC (FAT) \citep{fat}. Both of the pipelines fit the 3D titled ring models to the data cubes and have their own advantages and limitations. These are described in detail in the following sections.

We compare the fitting procedure and the results for each galaxy for these two pipelines. In the case of the BBarolo, we use the moment zero maps to mask the emission region in the data cube. In order to do that, we first use the task \say{BLANK} from AIPS with opcode \say{TVCU} on the moment zero map to select the region where the emission of the galaxy is seen. Then we again run the task \say{BALNK} on our data cube with opcode \say{IN2C} and the masked moment zero map as input. It creates a masked datacube where all the regions except the region where the emission from moment zero is seen are blanked. It becomes easier for BBarolo to deal with this datacube as the region of interest in the datacube is specified. However, BBarolo itself creates a channel-by-channel mask and fits the tilted-ring model.
The initial parameters used in the fitting of BBarolo are described below. The centres are given as the pixels where the optical centres lie. The ring size is given equal to the full-width half maxima (FWHM) of the major axis of the synthesized beam. The initial guesses of rotation velocity and dispersion velocity are the rough estimates anticipated from the moment one and moment two maps, respectively. And for the initial estimate of inclination and position angles, we have taken the values from \citet{biswas2022}, where inclination and PA are measured by fitting ellipse to moment zero map's 1 M$_\odot$pc$^{-2}$ contour. The code is first run keeping all the parameters except the systemic velocity free, i.e., the pixel positions of centre, rotation velocity, radial velocity, dispersion velocity, scale height, inclination angles and position angle. The systemic velocity is kept fixed and equal to zero  as the data cubes are made in the galaxy's reference frame \citep[see section 3 of][ for details]{biswas2022}. 

In the first run, we only take five rings for the fitting, and then the number of the rings is increased depending on produced moment maps from the data and model. After the first run, the centre gets converged to a fixed value and then on the consecutive runs, we keep the centre fixed on that value and vary all other mentioned parameters. However, for some galaxies, we have kept the scale height fixed to get a good fitting decided from the residuals of moment maps. The code is run meticulously and rigorously several times by changing the initial estimates depending upon the residual maps, PV diagram, and the ring-to-ring variations of different parameters for getting a successful fit. The code is run in two stages keeping the \say{TWOSTAGE} parameter in BBarolo to be \say{True} and considering emission from both the arms. The parameter minimised in the fitting procedure for some galaxies is $\chi^2$; for some galaxies, it is $|model-data|$. By evaluating fitting quality using residual moment maps and how well the fitting parameter matches in the two stages of fits, the parameter to be minimised in the fitting procedure is decided. Further, the asymmetric-drift correction is also being applied.

\begin{figure}
    \centering
    \includegraphics[height=11cm]{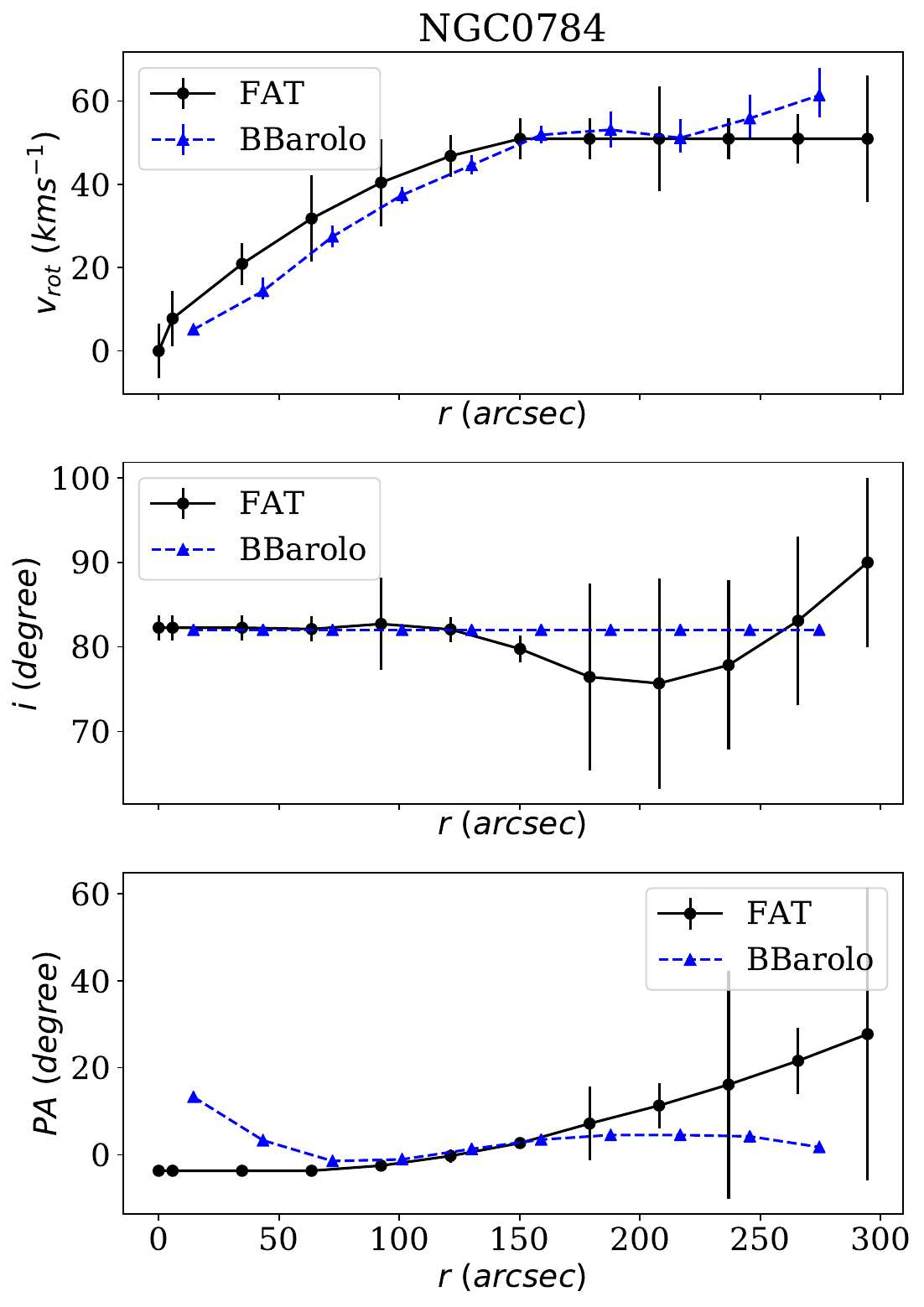} 
   \caption{Comparison of the rotation curve, inclination angle and position angle derived from FAT and BBarolo for NGC0784. }
    \label{fig:Comp_rotc_pa_inc}
\end{figure}

In the case of FAT, instead of masking the data cube with the moment zero maps, we remove some of the beginning and end channels of the cube to keep only the channels where the emission is seen. While preparing the data cubes, we already subtracted the continuum image from the UV data, and in some cases, if there was any residual continuum emission in the image cubes, the AIPS task IMLIN was performed to fit a first-order polynomial to some line-free channels and eliminate it from the image cube.  The initial guesses of different parameters are not required in the case of the FAT because it is a fully automated process. FAT, by default, uses the ring size equal to the FWHM of the major axis of the beam. In this way, the ring size used by both pipelines becomes equal. Instead of using the same ring size, the data points coming from both fittings appear to be a little shifted because BBarolo keeps the ring position at the centre of the beam while FAT keeps it at the inner edge of the ring. Figure \ref{fig:Comp_rotc_pa_inc} shows the comparison of the fitting of the rotation curve, inclination, and position angle with radius found from both the pipelines for the galaxy NGC0784; the comparison plots for all other sources can be viewed in
\href{http://www.physics.iisc.ac.in/~nroy/garcia_web/kinematics.html}{ GARCIA website, Kinematics section, figure 1 } \footnote{\label{garcia-kinematics}{\url{http://www.physics.iisc.ac.in/~nroy/garcia_web/kinematics.html}}} .

\begin{figure}
    \centering
        \includegraphics[height=7cm]{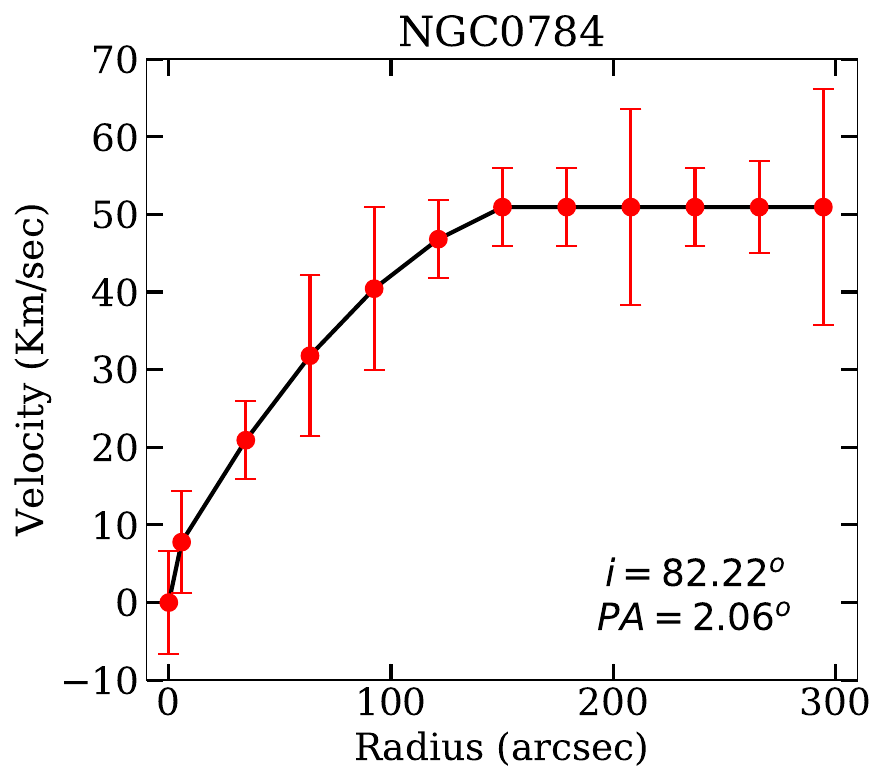}
        \caption{Rotation Curve of the galaxy NGC0784 derived using FAT.}
    \label{fig:fat_rotc}
\end{figure}

    After examining these rotation curves and their residuals and understanding the fitting procedure of these two pipelines, we decide to finalize the rotation curve found from FAT for most of the sources. The reasons behind it are mentioned below. For the purely resolved data, BBarolo does not work adequately for inclination angles $<45^0$ and $>75^0$. However, FAT works reasonably well in a wide range of inclination angles, from $20^0$ to $90^0$. When using the same ring size, FAT fits more number of rings (probing higher radius) to the rotation curve than BBarolo for all the cases. Most importantly, rotation curves may change significantly depending upon the initial guess of different parameters giving a successful fit for each case. On the other hand, FAT takes the initial estimate of the parameters from Source finding Application (\href{https://www.atnf.csiro.au/people/Tobias.Westmeier/tools_software_sofia.php}{ SoFiA }\footnote{\url{https://www.atnf.csiro.au/people/Tobias.Westmeier/tools_software_sofia.php}}, \citeauthor{sofia} \citeyear{sofia} )  and runs in a fully automated manner. It also has the capacity to fit structures like wraps or extra-planner gas.  For example, from the velocity field of NGC7741 \citep[][ figure 6]{biswas2022}, we see that there is a signature of wrap in the northeast side. The large inclination variations seen in FAT modelling suggest the presence of this wrap. However, BBarolo does not show any indication of wrap in its modelling.  
    The rotation curve from FAT for the galaxy NGC0784 is given in figure \ref{fig:fat_rotc}; the FAT-rotation curves for all sources are shown in  \href{http://www.physics.iisc.ac.in/~nroy/garcia_web/kinematics.html}{ the GARCIA website, Kinematics section, figure 2}. Besides that, to check the quality of the fitting, we examine the moment one maps residuals, i.e., the difference between the moment one map derived from the data and that derived from the model. In the case of the kinematic models obtained from FAT, we see that 90 to 99 percent of the data points of the moment one residual maps lie in between $\sim \pm$ 20 kms$^{-1}$ for all the sources except one, NGC4861. For this source, in the residual moment one map obtained from FAT, 90 percent of the data points lie in between $\sim \pm$ 33 kms$^{-1}$. However, in the moment one residual map obtained from BBarolo, 90 percent of the data points lie in between $\sim \pm$ 14 kms$^{-1}$. The moment one residual maps for the kinematic model obtained through FAT for all the sources and that of NGC4861 for the kinematic model obtained through BBarolo  are respectively shown in  figure 3 and 4b of the \href{http://www.physics.iisc.ac.in/~nroy/garcia_web/kinematics.html}{ Kinematic section of GARCIA website} .  

\begin{figure}
    \centering
        \includegraphics[height=7cm]{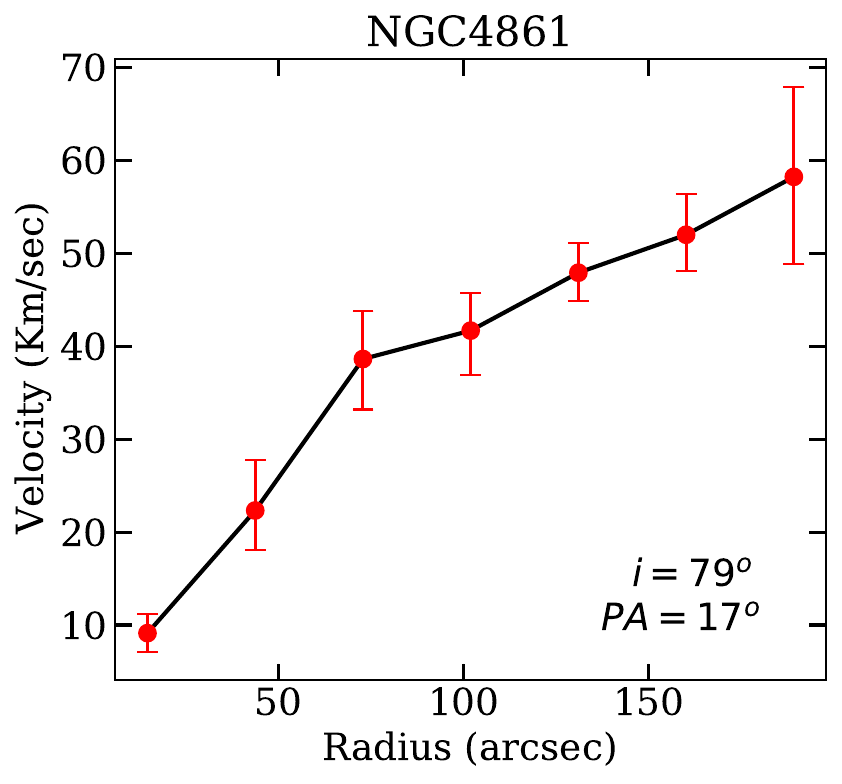}
        \caption{Rotation Curve of the galaxy NGC4861 derived using BBarolo.}
    \label{fig:fat_rotc_bb_ngc4861}
\end{figure}

    Out of these kinematic models of eleven galaxies, for ten galaxies, we selected kinematic modelling obtained through FAT and for these sources, all the results presented in this paper are derived from the FAT rotation curve. For one galaxy, NGC4861, which is an irregular dwarf galaxy and has a signature of the interaction with its close companion NGC4861B \citep{n4861_paper1}, we are unable to get a satisfactory fitting to the rotation curve obtained from FAT while doing the mass modelling. For this source, the rotation curve obtained using FAT gets flattened with a rotation velocity $V_{flat}$ $\sim$ 50 kms$^{-1}$ but the rotation curve obtained from BBarolo does not reach the flat part and keeps increasing with a rotation velocity  $V_{max}$ $\sim$ 70 kms$^{-1}$ which is more consistent with the maximum velocity inferred from the moment one map ($\ge$ 60 kms$^{-1}$, \citet[see figure 6,][]{biswas2022}). Besides that, for this source, we compared the moment one residual maps obtained in kinematic modelling using FAT and BBarolo and decided to finalize the rotation curve obtained from BBarolo. The finalized rotation curve for this source that is obtained from BBarolo is shown in figure \ref{fig:fat_rotc_bb_ngc4861} and as well as in figure  4a of \href{http://www.physics.iisc.ac.in/~nroy/garcia_web/kinematics}{the kinematics section of GARCIA website}. All the results shown in this paper, including the mass modelling, for this source, are done using the rotation curve obtained using BBarolo. 

\section{Comparison of inclination angle and the different measures of velocity}
\label{sec:comp_inc_vel}
\subsection{Inclination angle}
    Inclination angles play a vital role in determining the circular velocity of galaxies. We show the comparison of the inclination angles found from different methods in  table \ref{tab:comp_incl}. The optical inclinations (i$_{opt}$) given in table \ref{tab:comp_incl} are derived using the following way. In the procedure of finding the surface brightness of the stars through the Multi-Gaussian Expansion method, an ellipse is fitted considering the boundary of the galaxy.  The inclinations of the galaxies are found using the major axis ($a_{maj}$) and minor axis ($b_{minor}$) of that ellipse through the following equation \citep[see e.g.,][]{q0_eqn}: 
    \begin{equation}
       \sin(i_{opt})= \sqrt{\frac{1-(b_{minor}/a_{maj})^2} {1-q_o^2}} ,  
     \end{equation}
      where $q_o$ is the intrinsic axial ratio: for spirals, $q_o \sim 0.2$ \citep[see e.g.,][]{q0_spirals, q0_spirals2} and for dwarfs, $q_o \sim 0.6$ \citep{q0_dwarf1}. 
    
     The H~{\sc i} inclination angle (i$_{mom0}$) is derived in a similar way, i.e., from the ratio of the major axis (a$_{HI}$) to the minor axis (b$_{HI}$) of the fitted ellipse to our derived moment zero map's surface density ($\Sigma_{HI}$) of $1$ M$_{\sun}$\thinspace pc$^{-2}$.
    
\begin{figure}
    \centering
    \includegraphics[height=6.3cm]{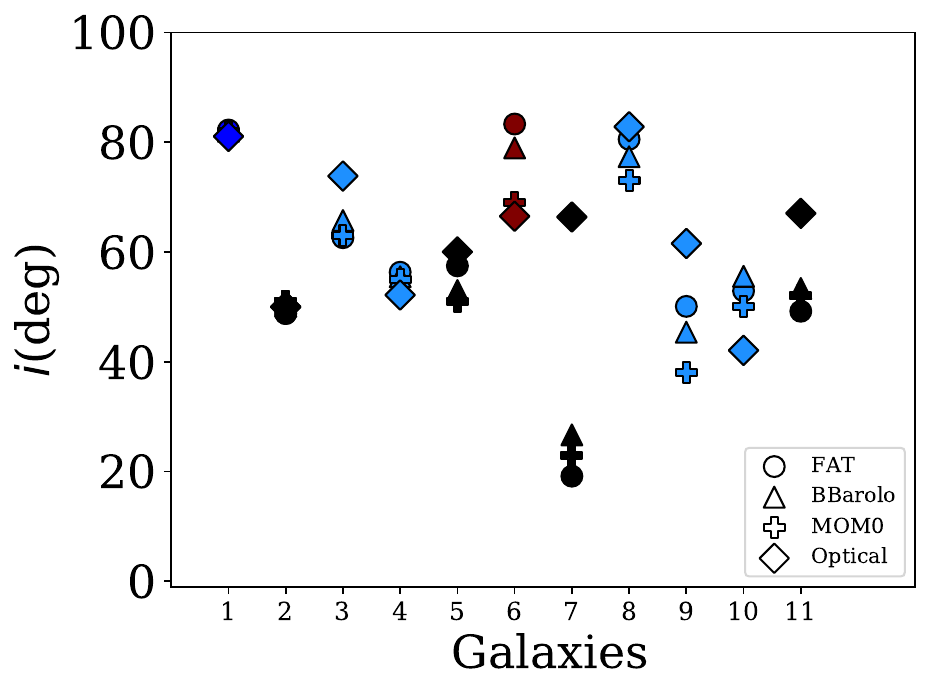}
    \caption{Comparison of inclination angle found from different methods. Galaxies are colour-coded according to their Hubble type;  Sc: Dodgerblue, Sd: Blue, Sm: Maroon, Ir: Black.  }
    \label{fig:comp_incl}
\end{figure}

    In table \ref{tab:comp_incl}, i$_{BBarolo}$ and i$_{FAT}$ respectively denote the kinematic inclination angle found from BBarolo and FAT. For each galaxy, as the inclination angle varies from ring to ring, we take the median value as the inclination angle of the galaxy. Table \ref{tab:comp_incl} shows the median values of the inclination angles for each galaxy. Figure \ref{fig:comp_incl} shows a comparison of i$_{opt}$, i$_{mom0}$, i$_{BBarolo}$ with and i$_{FAT}$. From the figure, we see that for most of the cases, i$_{mom0}$, i$_{BBarolo}$, and i$_{FAT}$ give similar values. For some cases, e.g., for NGC7292 and NGC7800,  i$_{opt}$ differs significantly from each three of them. This large difference between kinematic inclination and optical inclination can cause significant changes  in the deprojected velocity determined from the widths of the global H~{\sc i} spectra and hence the Baryonic Tully-Fisher relation.

\begin{table}
    \centering
   \begin{tabular}{ccccc}
     \hline
      Source & $i_{opt}$ & $i_{mom0}$ & $i_{BBarolo}$ & $i_{FAT}$   \\
      name & (deg) & (deg) & (deg) & (deg) \\
      \hline
	NGC0784 & 81.1 & 82.0 & 82.0 & 82.22 \\
	NGC1156 & 50.0 & 51.0 & 50.15 & 48.79 \\
	NGC3027 & 73.8 & 63.0 & 65.72 & 62.6 \\
	NGC3359 & 52.2 & 55.0 & 55.48 & 56.31 \\
	NGC4068 & 60.0 & 51.0 & 52.98 & 57.47 \\
	NGC4861 & 66.5 & 69.0 & 78.96 & 83.31 \\
	NGC7292 & 66.4 & 23.0 & 26.68 & 19.16 \\
	NGC7497 & 82.8 & 73.0 & 77.35 & 80.53 \\
	NGC7610 & 61.6 & 38.0 & 45.39 & 50.09 \\
	NGC7741 & 42.1 & 50.0 & 55.56 & 52.91 \\
	NGC7800 & 67.0 & 52.0 & 53.35 & 49.2 \\
\hline
\end{tabular} 
    \caption{Comparison of the inclination angles derived from the various methods. }
    \label{tab:comp_incl}
\end{table}

\subsection{Velocity}
As mentioned in the introduction, different velocity definitions have been used in literature to construct the Baryonic Tully-Fisher relation (BTFR). \citet{BFTR1} showed that $V_{flat}$ gives the tightest BTFR. Their study suggests the importance of establishing the BTFR using interferometric observation. However, the H~{\sc i} rotation curves used by \citet{BFTR1} are mostly derived by fitting 2D titled ring model to the velocity field. 

Here in this study, we find out $V_{flat}$ and $V_{max}$ from the rotation curve derived by fitting the 3D tilted ring model through FAT. The rotation curve for all of our sources obtained through FAT reaches the flat part, and $V_{flat}$ is measured as the average velocity in the flat part. In the case of the NGC4861, where the rotation curve from BBarolo does not reach the flat part, we take the maximum velocity as equal to $V_{flat}$. And, $V_{max}$ is the maximum velocity found in the rotation curve. We compare these velocities with velocity widths found from global H~{\sc i} spectra from single-dish observation and interferometric observation. The single-dish spectra are taken from \say{A Digital Archive of H I 21 Centimeter Line Spectra of Optically Targeted Galaxies} \citep{sngl_dish1} and \say{The Arecibo Legacy Fast ALFA Survey: The ALFALFA Extragalactic H I Source Catalog} \citep{sngl_dish2}. The H~{\sc i} interferometric spectra are taken from \citet{biswas2022}. Table \ref{tab:comp_incl} shows the comparison of different measures of velocities.

\begin{figure*}
    \centering
     \begin{subfigure}[b]{0.4\textwidth}
         \centering
         \includegraphics[width=\textwidth]{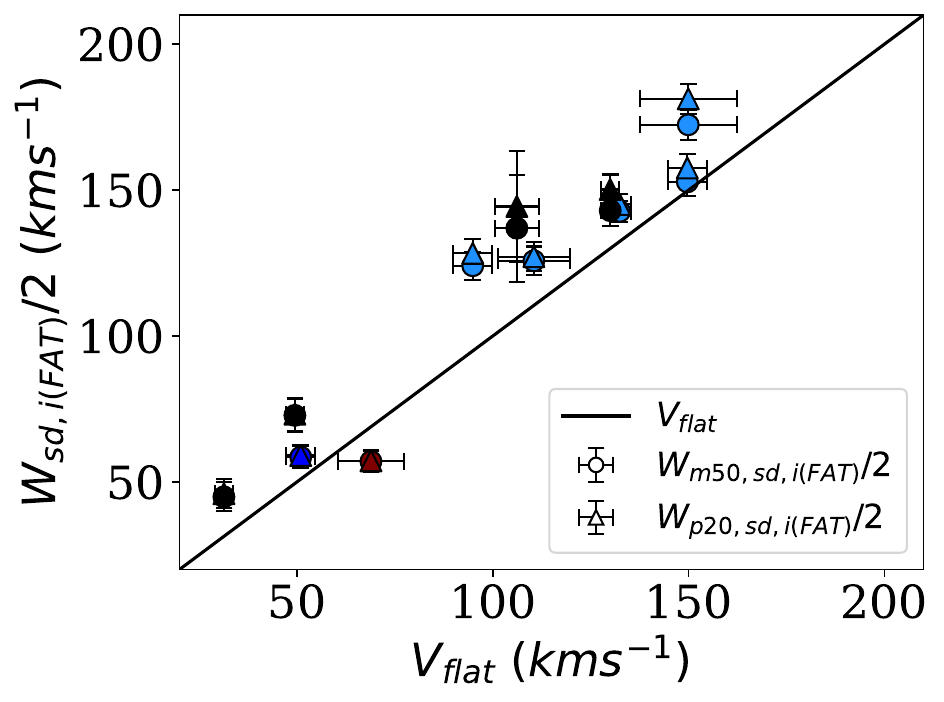}
         \caption{}
         \label{fig:vflat_wsd_icor}
     \end{subfigure}
     \begin{subfigure}[b]{0.4\textwidth}
         \centering
         \includegraphics[width=\textwidth]{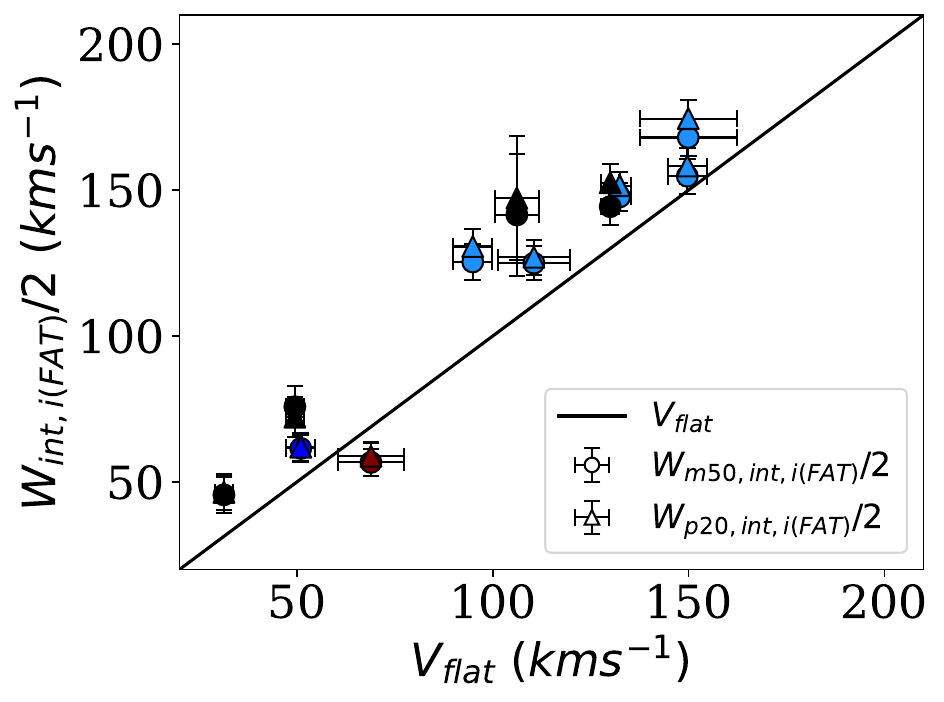} 
         \caption{}
         \label{fig:vflat_wint_icor}
     \end{subfigure}
     \begin{subfigure}[b]{0.4\textwidth}
         \centering
         \includegraphics[width=\textwidth]{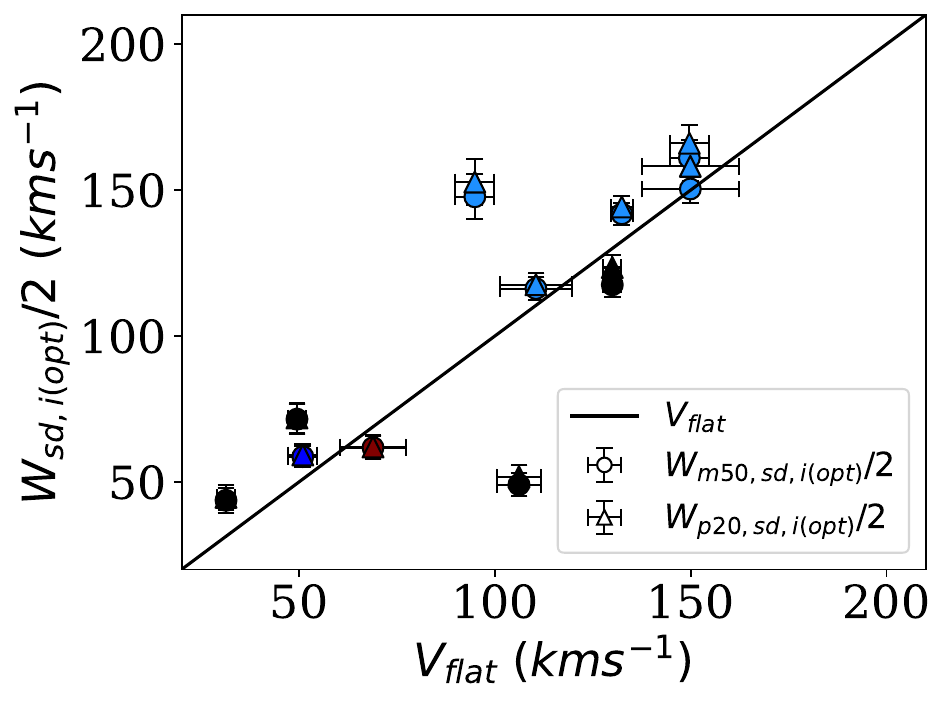}
         \caption{}
         \label{fig:vflat_wsd_opt_icor}
     \end{subfigure}
      \begin{subfigure}[b]{0.4\textwidth}
         \centering
         \includegraphics[width=\textwidth]{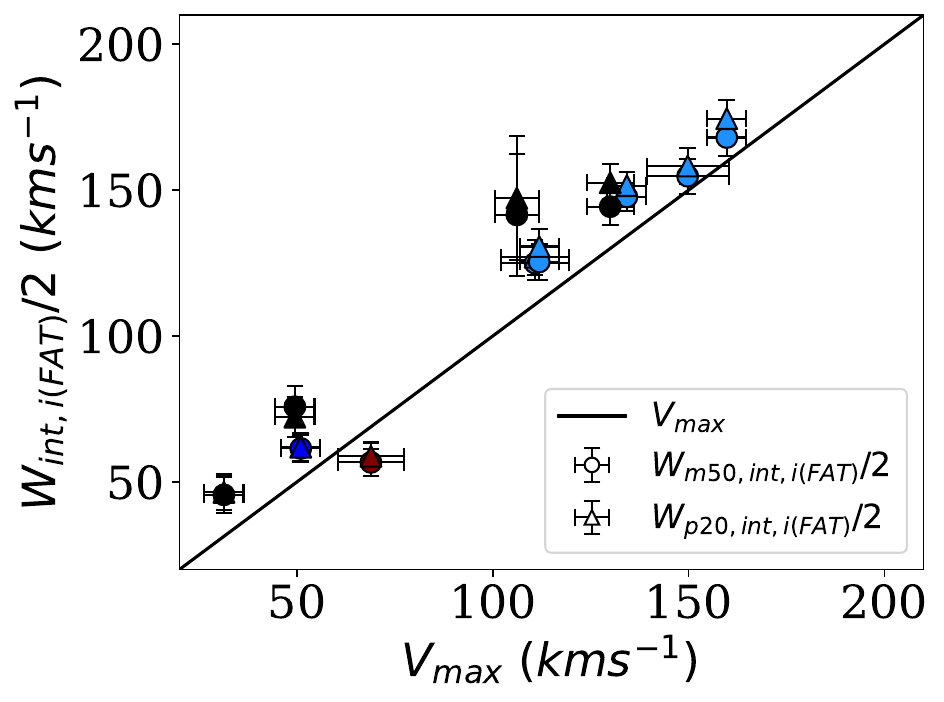}
         \caption{}
         \label{fig:vmax_w_int_icor}
     \end{subfigure}
    \caption{ Figure \ref{fig:vflat_wsd_icor} shows the comparison of the velocity widths ($W_{p20}/2$ and $W_{p50}/2$) derived from the single-dish H~{\sc i} spectra corrected for the kinematic inclination angle $i_{FAT}$ with $V_{flat}$. Figure \ref{fig:vflat_wint_icor} shows the comparison of the velocity widths ($W_{p20}/2$ and $W_{p50}/2$) derived from the interferometric global H~{\sc i} spectra corrected for the kinematic inclination angle $i_{FAT}$ with $V_{flat}$. Figure \ref{fig:vflat_wsd_opt_icor}  shows the comparison of the velocity widths ($W_{p20}/2$ and $W_{p50}/2$) derived from the single dish global H~{\sc i} spectra corrected for the  optical inclination angle $i_{opt}$ with $V_{flat}$ . Figure \ref{fig:vmax_w_int_icor}  shows the comparison of the velocity widths ($W_{p20}/2$ and $W_{p50}/2$) derived from the interferometric global H~{\sc i} spectra corrected for the  kinematic inclination angle $i_{FAT}$ with $V_{max}$.
    The black solid line in figure \ref{fig:vflat_wsd_icor}, \ref{fig:vflat_wsd_icor}, \ref{fig:vflat_wsd_opt_icor} and \ref{fig:vmax_w_int_icor}  respectively represent $V_{flat}$ = $W_{sd,i(FAT)}/2$, $V_{flat}$ = $W_{int,i(FAT)}/2$, $V_{flat}$ = $W_{sd,i(opt)}/2$ and $V_{max}$ = $W_{int,i(FAT)}/2$. Galaxies are colour-coded according to their Hubble type, mentioned in figure \ref{fig:comp_incl}. }
    \label{fig:comp_vel}
\end{figure*}

We compare the systemic difference between $V_{flat}$ and $V_{max}$  with the rotation velocities of the galaxies defined from the width of the single-dish spectra ($W_{p20,sd}$ and ($W_{m50,sd}$) and interferometric H~{\sc i} spectra ($W_{p20,int}$) and ($W_{m50,int}$).  We correct the widths of the single-dish spectra with both the kinematic inclination  ($i_{FAT}$) and optical inclination ($i_{opt}$); interferometric spectra for the kinematic inclination  ($i_{FAT}$) and  compare them with $V_{flat}$.  Figure \ref{fig:comp_vel} shows the results of the comparison. From Figure \ref{fig:vflat_wsd_icor} and \ref{fig:vflat_wint_icor}, we see that both the single-dish and interferometric spectra corrected for kinematic inclination have similar velocity, but both of them has an average higher value from $V_{flat}$. Nevertheless, figure \ref{fig:vflat_wsd_opt_icor} shows that the single-dish spectra corrected for the optical inclination have little higher scatter than those corrected for kinematic inclination. This result suggests the importance of correcting the H~{\sc i} spectra with the kinematic inclination instead of the optical inclination. Further, figure \ref{fig:vmax_w_int_icor} shows that kinematic inclination corrected H~{\sc i} spectra matches little better with $V_{max}$ than $V_{flat}$. This is perhaps because the width of the H~{\sc i} spectra traces the maximum projected rotation velocity, and it can correlate with the maximum circular velocity of the galaxy instead of the rotation velocity where the rotation curve gets flat. This may also be the reason for the average higher value of inclination corrected widths in comparison to $V_{flat}$.

\begin{table}
    \centering
    \setlength{\tabcolsep}{2pt} 
    \begin{tabular}{ccccccc}
    \hline
    Source & W$_{m50}$,int & W$_{m50}$,sd & W$_{p20}$,int & W$_{p20}$,sd & V$_{flat}$ & V$_{max}$ \\
    name & (kms$^{-1}$) & (kms$^{-1}$) & (kms$^{-1}$) & (kms$^{-1}$) & (kms$^{-1}$) & (kms$^{-1}$) \\
    \hline
        NGC0784 & 60.9 & 58.0 & 61.3 & 58.5 & 50.9 & 50.9 \\
        NGC1156 & 57.0 & 54.8 & 54.3 & 55.1 & 49.4 & 49.4 \\
        NGC3027 & 110.9 & 111.6 & 112.8 & 112.9 & 110.5 & 110.8 \\
        NGC3359 & 128.8 & 127.1 & 131.7 & 131.1 & 149.7 & 149.7 \\
        NGC4068 & 38.4 & 37.8 & 39.1 & 38.7 & 31.3 & 31.3 \\
        NGC4861 & 56.3 & 56.6 & 58.4 & 56.8 & 68.9 & 68.9 \\
        NGC7292 & 46.4 & 44.9 & 48.3 & 47.4 & 106.1 & 106.1 \\
        NGC7497 & 145.7 & 140.8 & 149.5 & 143.1 & 132.4 & 134.2 \\
        NGC7610 & 128.9 & 132.2 & 133.9 & 139.1 & 149.9 & 159.7 \\
        NGC7741 & 100.0 & 99.0 & 104.2 & 102.4 & 94.9 & 118.0 \\
        NGC7800 & 109.3 & 108.2 & 115.5 & 113.7 & 130.0 & 130.0 \\
  \hline
    \end{tabular}
    \caption{Comparison of different measures of velocities}
    \label{tab:comp_vel}
\end{table}

\section{Mass modeling}
\label{massmodel}
The total velocity of the galaxy can be decomposed into velocities due to luminous matter and dark matter halo, where the luminous matter considers the velocities coming due to stellar disk and gas disk. So, the total velocity can be modelled as the following equation:

\begin{equation}
(V^{tot}_{mod})^2 = V_{gas}^2 + V_{star}^2 + V_{halo}^2
\end{equation}

Where $V_{gas}^2$, $V_{star}^2$, $V_{halo}^2$ are respectively the velocity component due to gas, star and dark matter halo.

\subsection{Stellar component}
The most common way to measure the galaxies' surface brightness or luminosity distribution is by isophot fitting. However, a strong deviation in the isophote from the ellipse can not be modelled properly in this ellipse fitting method. Thus, for multi-component objects such as lenticular galaxies and spiral galaxies, which have a bulge and a disk or a nuclear disk or bar, this method of ellipse fitting is not well suited for photometric modelling. Hence, to determine the luminosity distribution of these galaxies, we follow the multi-Gaussian expansion (MGE) method \citep{mge2, mge3, mge_cappellari2002}, which overcomes these limitations.  In the MGE method, a series expansion of the luminosity distribution is performed using two-dimensional Gaussian functions. The projected surface brightness ($\Sigma_{star}$) obtained from the MGE method can be written as following equations \citep[see][section 2]{mge_cappellari2002}:

\begin{align}
\label{mge}
    & \Sigma_{star} (R, \theta)  = \sum_{i=1}^N \frac{L_i}{2 \pi \sigma_i^2 q_i} exp \Bigl[ - \frac{1}{2\sigma_i^2} \Bigl( x_i^2 + \frac{y_i^2}{q_i^2} \Bigr)   \Bigr], \\
    & x_i  = Rsin(\theta - \Psi_i) \nonumber \\
    & y_i   = Rcos(\theta - \Psi_i) \nonumber
\end{align}

where $\Sigma_{star}$ is written in the polar coordinates on the sky plane $(x,y)$. N is the number of Gaussian components; $L_i$ is the total luminosity, $q_i$ is the observed axial ratio, $\sigma_i$ is the dispersion along the major axis, and $\Psi_i$  is the position angle of each Gaussian. As mentioned in the section \ref{data} we used  both the optical and infrared data to compute the stellar contribution in the total velocity. For both cases, we first mask the nearby brightest objects in the background or the foreground. Then following the MGE procedure, find out their luminosity distribution. 

To find the stellar kinematics from the results of the MGE fitting, we use the Jeans anisotropic multi-Gaussian expansion (JAM; \citet{jam_cappellari_2008}) model. The stellar dynamical equation that describes the stellar circular velocity curves (CVC) is given by the following equation:

\begin{align}
  \begin{split}
    V_{star}^2 (R) = \sum_{i=0}^N \frac{2GL_i(M/L)}{\sqrt{2\pi}\sigma_i} \frac{R^2}{\sigma_i^2} \int_0^1 exp\Bigl( -\frac{u^2R^2}{2\sigma_i^2} \Bigr) \times \\ \frac{u^2du}{\sqrt{1-(1-q_i^2)u^2}} ,
  \end{split}
\end{align}

where $L_i$, $\sigma_i$, $q_i$ are parameters of each Gaussian as defined above.  Here, we have considered the mass-to-light ratio (M/L) to be constant  with the radius for this model.

\subsection{Gas component}
\label{gas_compnt}
To find out the gas component of the velocity, we used the surface density found from 3D kinematic modelling of the H~{\sc i} interferometric data through FAT. The surface densities are further multiplied by a factor of 1.33 to account for the He contribution. The velocity of gas given the surface density can be found through the following equation \citep[see][equaion 2.188]{gal_dyn_bt}:

\begin{equation}
V_{gas}^2 (R) = 2\pi G R \int_0^{\infty} dk k J_1(kR) \int_0^{\infty} dR^{\prime} R^{\prime} \Sigma(R^{\prime}) J_0(k R^{\prime})
\label{eqn:vgas}
\end{equation}

Where $\Sigma$ is the surface density and $J_0$ and $J_1$ are the Bessel's functions. The integration can be thought to be as two Hankel transformations. The integration over $R^{\prime}$  is the Hankel transformation of the surface density $\Sigma(R^{\prime})$, and the second one is the Hankel transformation of the result found from the first integration. We make use of the Python package \href{https://hankl.readthedocs.io/en/latest/}{Hankl} \footnote{\url{https://hankl.readthedocs.io/en/latest/}} \citep{hankl} to compute the integration. This code performs the Hankel transformation considering the oscillatory nature of the integrands. Following the equation \ref{eqn:vgas}, we can associate the error in V$_{gas}$ as follows:

\begin{equation}
    \frac{\delta V_{gas}(R)}{V_{gas}(R)} = \frac{1}{2} \frac{\delta \Sigma}{\Sigma}
\end{equation}

Where, the $\delta \Sigma$ is error in $\Sigma$ and defined by the following equation:
\begin{equation}
     \delta \Sigma =  \frac{17.79}{\nu_0^{2} \theta_1 \theta_2} (\delta M_0)
\end{equation}
where $ \Sigma$ is in M$_{\sun}$pc$^{-2}$; $\nu_0$ is the rest frequency of H~{\sc i} in GHz; $\theta_1$, $\theta_2$ are the synthesized beam sizes in arc second; and M$_0$ is the moment zero map in mJy/beam-km/s unit. A representative error in the moment zero map can be defined as the multiplication of three times the RMS noise of a line-free channel of the H~{\sc i} data cube and the median of the gas dispersion velocity from the moment two maps. The accuracy of this process is tested and mentioned in detail in appendix \ref{vgas_accr}. 

\subsection{Dark matter halo profile}
As mentioned in the introduction, for the dark matter distribution, we are using the  NFW density profile \citep{navarro_1996}  given by the following equation:

\begin{equation}
    \rho(r) = \frac{\rho_0}{(\frac{r}{r_s})(1+\frac{r}{r_s})^2} ,
\end{equation}

Where $\rho_0$ is the central density, and $r_s$ is the scale radius. The velocity that corresponds to this density profile is given by the following equation:

\begin{align}
\begin{split}
        V_{halo} (R) = 0.014 M_{200}^{1/3} &\sqrt{\frac{20.24M_{200}^{1/3}}{R}}\times \\  
        & \sqrt{\frac {\ln (1+ \frac{RC}{20.24M_{200}^{1/3}}) - \frac{RC/(20.24M_{200}^{1/3})}{1+RC/(20.24M_{200}^{1/3)}} } {\ln(1+C) - \frac{C}{1+C}} },
    \end{split}
\end{align}

Where $M_{200}$ is the average halo mass enclosed in a sphere with a density 200 times  higher than the critical density of the Universe, and $C$ is the concentration index of the profile. To derive this equation, we have made use of the following relation: V$_{200}$ = 10H$_{0}$R$_{200}$, where H$_{0}$ = 72 kms$^{-1}$ Mpc$^{-1}$ \citep{Komatsu2009}. The halo mass (M$_{halo}$) is referred to as M$_{200}$ in the rest of the paper. 

As mentioned in the introduction, to fit the total velocity, $V_{tot}$, we use Markov Chain Monte Carlo (MCMC) sampler \citep[e.g.,][]{sivia1998} to locate the optimized values of the parameters of our model. The M/L, M$_{200}$ and C are our model's free parameters/priors. Although M/L can vary radially and considering the radial variation of M/L can better constrain the model, here in this study, we consider a simpler case by assuming it to be constant with radius. We use the code   developed by \cite{Tyulneva2021} to implement the fitting process \citep[also see][for similar MCMC code]{kalinova2017a} . The MCMC process implemented in this code is via the python package 'emcee' by \citet{emcee}. It is an affine-invariant ensemble sampler for MCMC. In this an inherently Bayesian fitting process; we used an exponential likelihood, $exp(-\frac{1}{2}\chi^2)$; where the $\chi$ is given by:

\begin{equation}
    \chi^2 =  \sum_{i=0}^N \Bigl[\frac{V_{tot,i} - V_{mod,i}(r_i,\theta)}{\sigma_{V_{tot,i}}} \Bigr]^2
\end{equation}

Where $i$ stands for radius; $V{tot,i}$ and $\sigma_{V_{tot,i}}$ are the total velocity and error in total velocity of the $i$th radius, respectively, and $\theta$ is a set of parameters to be fitted for each model. 

In most of the previous studies of mass modelling using MCMC optimization method, to select the priors for M$_{200}$ and $C$, different groups have made use of the M$_{star}$- M$_{200}$ relation \citep{moster2013} or/and M$_{200}$-$C$ relation \citep{maccio2008, dutton2014}. For example, \citet{sparc_mass_model} have used both of these relations and also flat priors separately for modelling the rotation curve and estimation of the distribution of the dark matter halo. They pointed out that the fitting quality decreases for priors based on M$_{star}$- M$_{200}$ and M$_{200}$-$C$ relation in comparison to flat priors. However, the  M$_{star}$- M$_{200}$ relation gets better matched when priors based on M$_{star}$- M$_{200}$ and M$_{200}$-$C$ relation, is used.  \citet{Teodoro2022_3Dmass_model_mcmc} got successful fits while using a flat priors for M$_{200}$ and lognormal prior for $C$ based on the  M$_{200}$-$C$ relation \citep{dutton2014}. \citet{massmod_pina2022} took flat priors for  M$_{200}$. For $C$, they separately examined two cases, first with flat distribution and secondly a Gaussian prior based on M$_{200}$-$C$ relation. They found that for some galaxies, flat prior in $C$ can not constrain its value well. However, all the free parameters are well constrained while using a Gaussian prior on $C$  based on the   M$_{200}$-$C$ relation. Here, in this study, we have used such a  set of priors for M$_{200}$ and $C$ that neither depends on  M$_{star}$- M$_{200}$ nor on M$_{200}$-$C$ relation and still provides successful fits and acceptable posterior distribution of the free parameters. The derived parameters from mass modelling are also being  compared with the existing scaling relations for consistency. These have been discussed in detail in the next sections. In figure 16 of \citet{dutton2014}, we see that observation from DiskMass survey \citep{diskmass2010, diskmass2013} suggests that spiral galaxies have halo mass, M$_{200}$ $\sim$  $10^{11.225}$ M$_{\sun}$ and concentration parameter, $C$ $\sim$ $10$. Based on this figure, we have set a lognormal prior for both M$_{200}$ and $C$. For M$_{200}$,  the mean and scale of the lognormal distribution are  $10^{11.225}$ and $0.5$ dex around this value. For $C$, the mean and the scale of the lognormal distribution are $10$ and $5$, respectively. In addition to that, to limit the infinite spread of the lognormal distribution, we have also imposed a uniform prior distribution in a broad range: for  M$_{200}$, it is within [0.01, 10$^{20}$] and for $C$, it is within [0.01, 1000]. However, the distribution of the starting points of the walkers is uniform but within a smaller range of M$_{200}$ (within [10$^{6}$, 10$^{15}$]) and $C$ (within [1, 45], these limits are loosely based on the values corresponding to M$_{200}$ $\sim$ 10$^{6}$ M$_{\sun}$ and M$_{200}$ $\sim$ 10$^{15}$ M$_{\sun}$ from M$_{200}$-$C$ relation \citep{maccio2008} ).  

As mentioned previously, the stellar population synthesis model and colour-magnitude diagram suggest that for all galaxies, M/L is tightest in 3.6 $\mu m$ in comparison to other wavelengths,  and  its value is nearly equal to 0.5. \citet{sparc_mass_model} used lognormal distribution in M/L in prior centred at 0.5 with a standard deviation of 0.1 for disk and centred at 0.7 with a standard deviation of 0.1 for the bulge for the mass modelling with 3.6-micron data. Thus for the 3.6-micron data, we further use a lognormal distribution in prior to M/L centred at 0.6 and with a standard deviation of 0.2. In addition to that, we also impose a uniform distribution on M/L within the range [0.01, 100] to limit the infinite spread of the lognormal distribution. However, the starting points of walkers are distributed uniformly in the range of [0.01, 3]. While using the r-band data, we have used a uniform distribution of M/L within the range of [0.01, 10]. The distribution of the starting points of the walkers also follows the same uniform distribution. 
For all the galaxies, either using 3.6 $\mu m$ data or r-band data, we have used $130$ walkers, $20000$ burning steps and $30000$ posterior steps.

\begin{figure*}
    \centering
    \begin{tabular}{cc}
          \includegraphics[height=5cm]{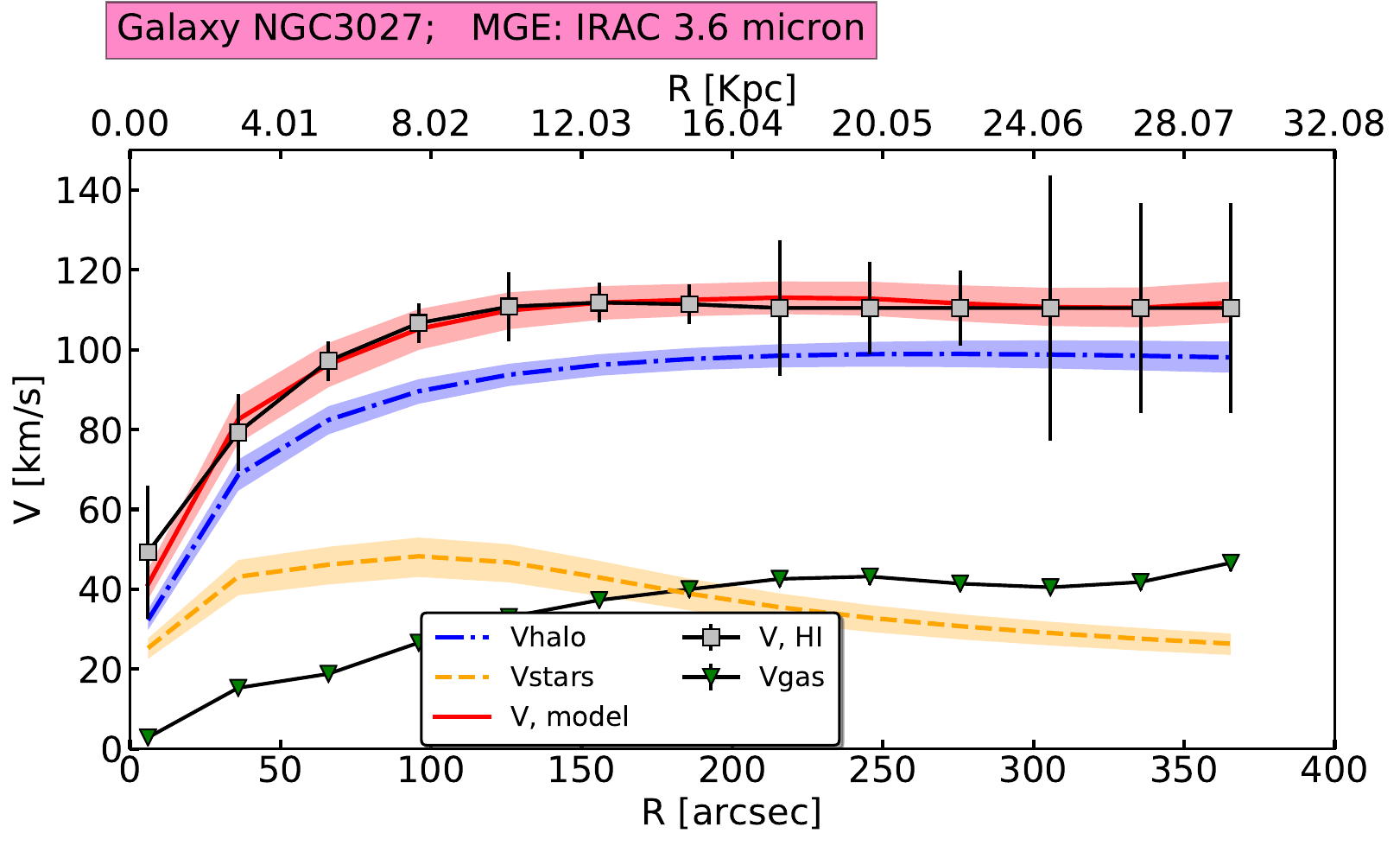} &
          \includegraphics[height=5cm]{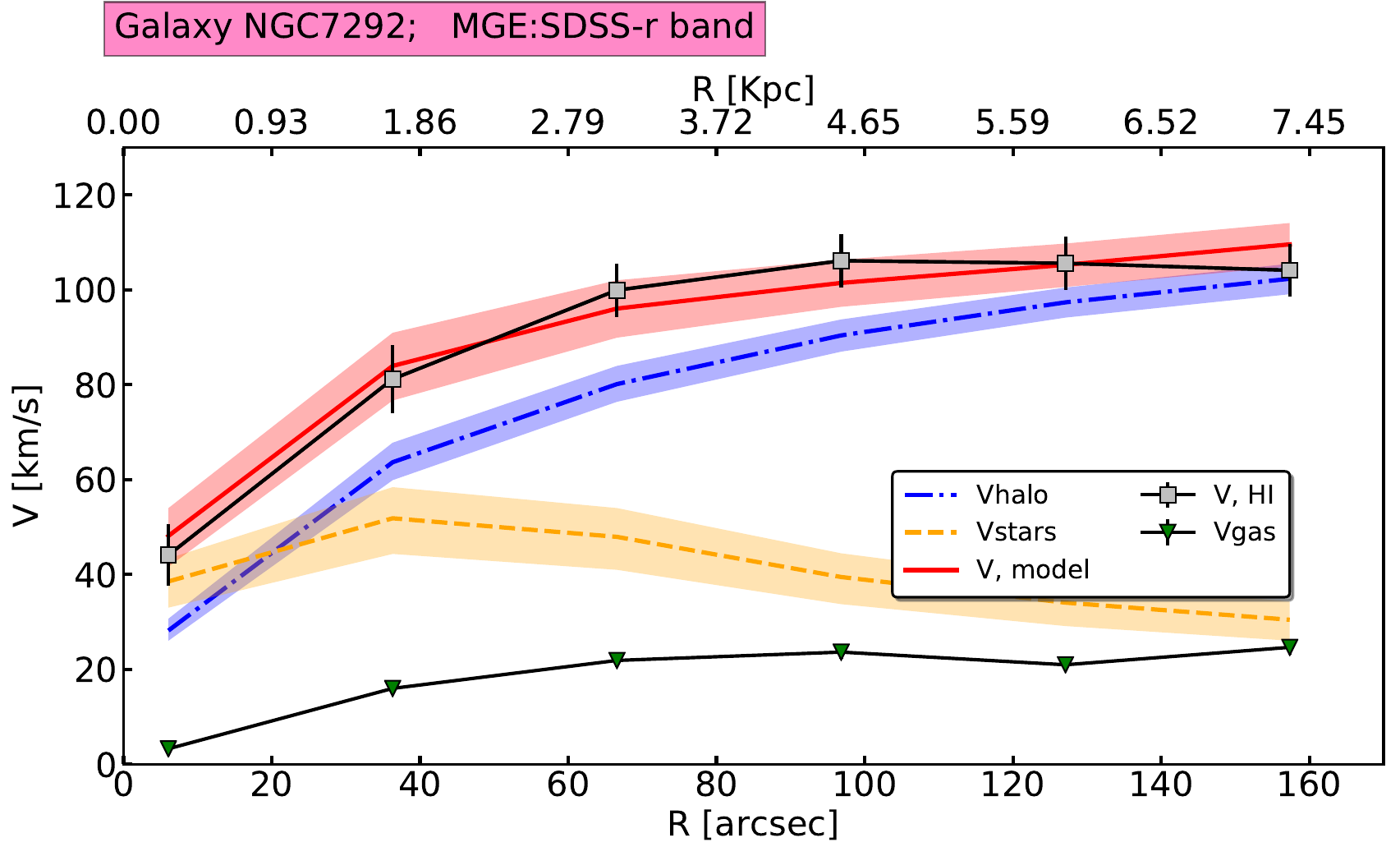} \\
          \includegraphics[height=7cm]{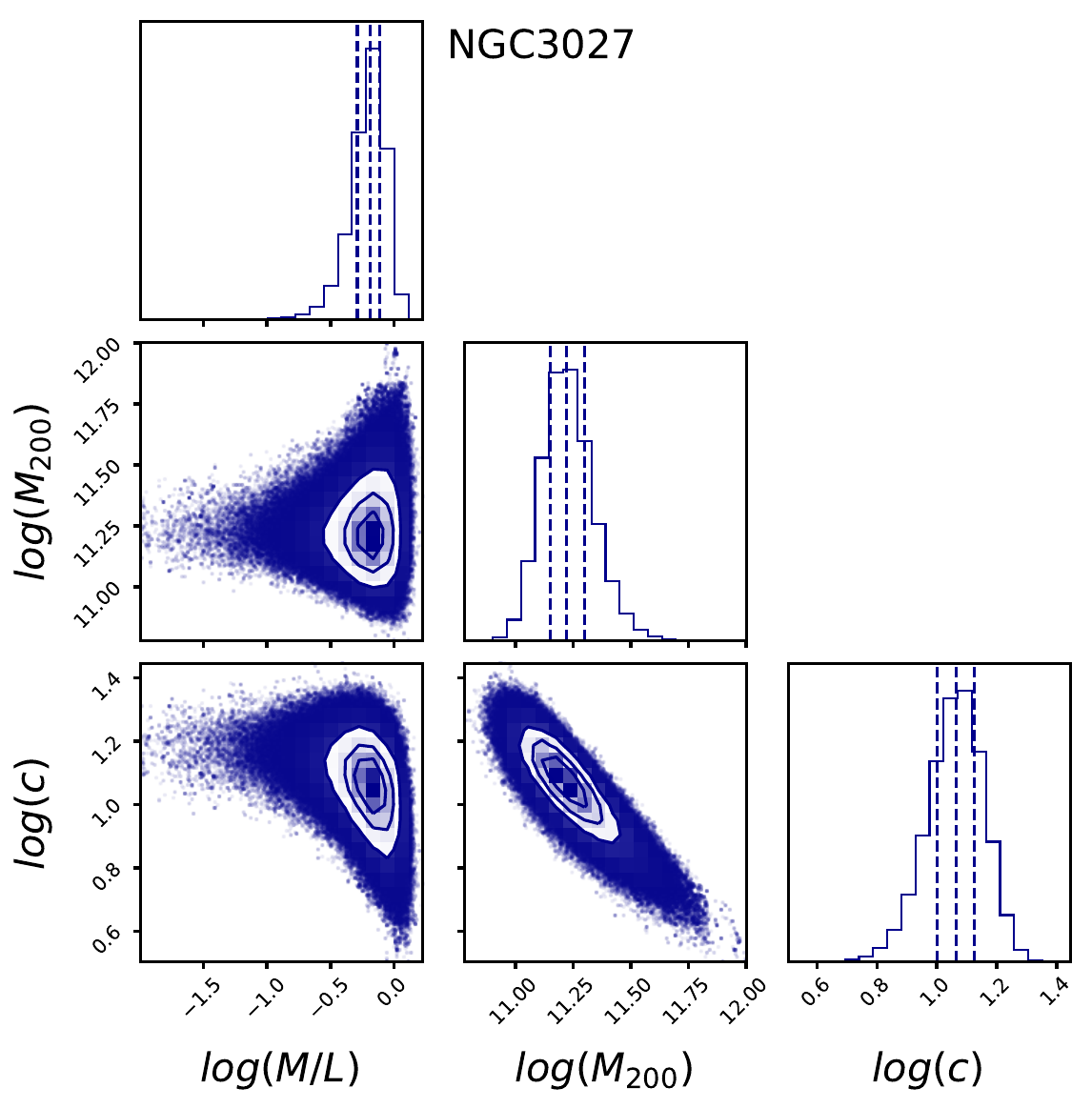} &
          \includegraphics[height=7cm]{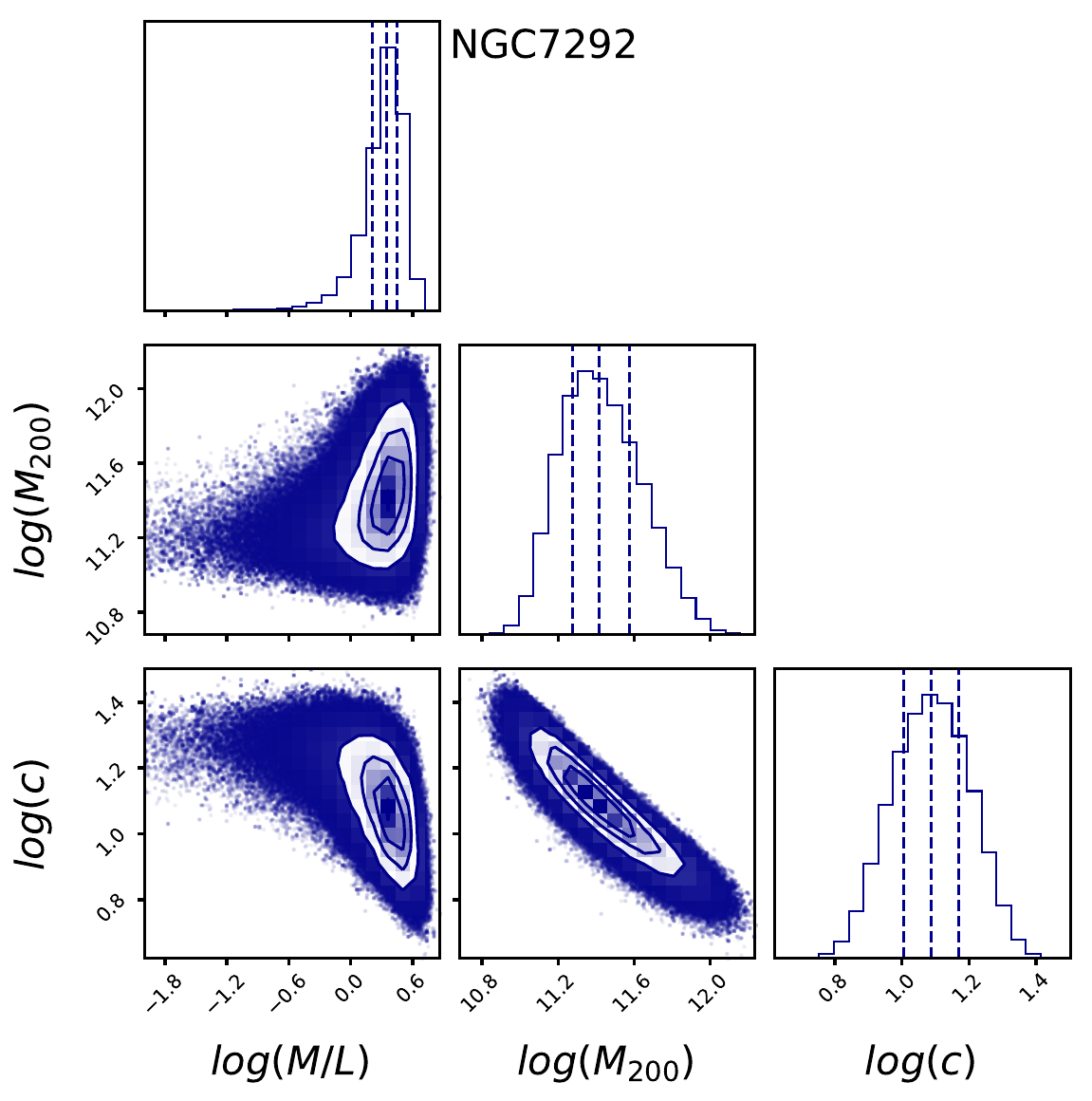} \\
    \end{tabular}
    \caption{  Modelled rotation curve along with the contribution of star, gas and halo for the source NGC3027 using 3.6 $\mu m$ data and for source NGC7292 using r-band data. The posterior distribution of different parameters (M/L, M$_200$ and C) used for modelling for each case is shown below each of the modelled rotation curves. }
    \label{fig:mass_model}
\end{figure*}

As mentioned previously, we fitted the model of total velocity for stellar dynamics found from different data sets separately. Figure \ref{fig:mass_model} shows the mass modelling of two galaxies from our results: NGC3027 using 3.6-micron data and NGC7292 using  r-band data. The posterior distribution of the fitted parameters for each case of the modelling is shown just below the modelled rotation curves. The similar plots for all the galaxies are presented in figure 1 of the
 \href{http://www.physics.iisc.ac.in/~nroy/garcia_web/mass_model.html}{ Mass models section of the GARCIA website}\footnote{\url{http://www.physics.iisc.ac.in/~nroy/garcia_web/mass_model.html}}. Further, 
 table \ref{tab:chisq} demonstrates the quality of the fitting in terms of reduced $\chi^2$ for each of the galaxies. 

\begin{figure*}
    \centering
    \begin{subfigure}[b]{0.4\textwidth}
         \centering
         \includegraphics[width=\textwidth]{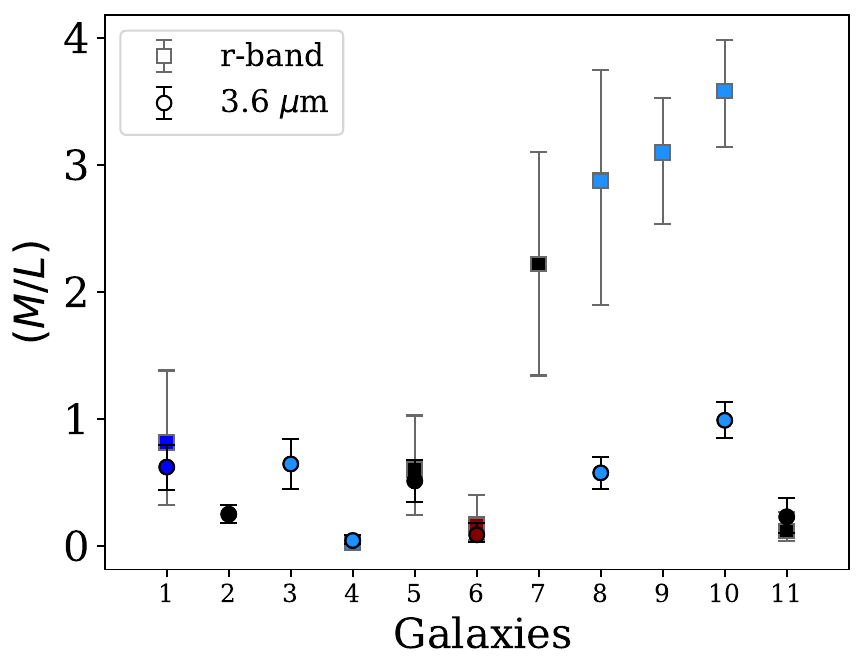}
         \caption{}
         \label{fig:mcmc_ml}
     \end{subfigure}
         \begin{subfigure}[b]{0.4\textwidth}
         \centering
         \includegraphics[width=\textwidth]{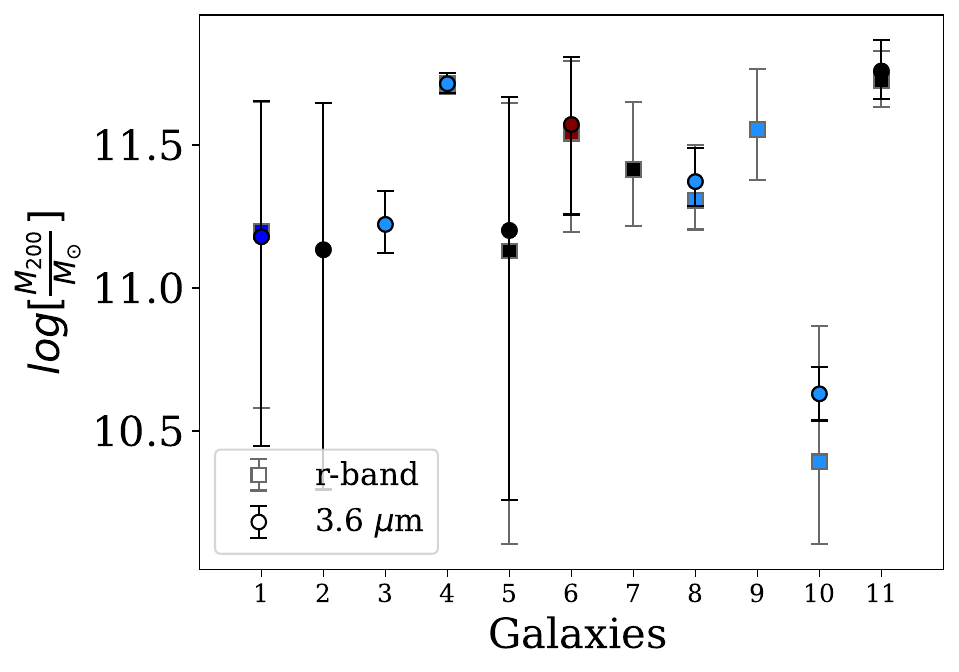}
         \caption{}
         \label{fig:mcmc_m200}
     \end{subfigure}
         \begin{subfigure}[b]{0.4\textwidth}
         \centering
         \includegraphics[width=\textwidth]{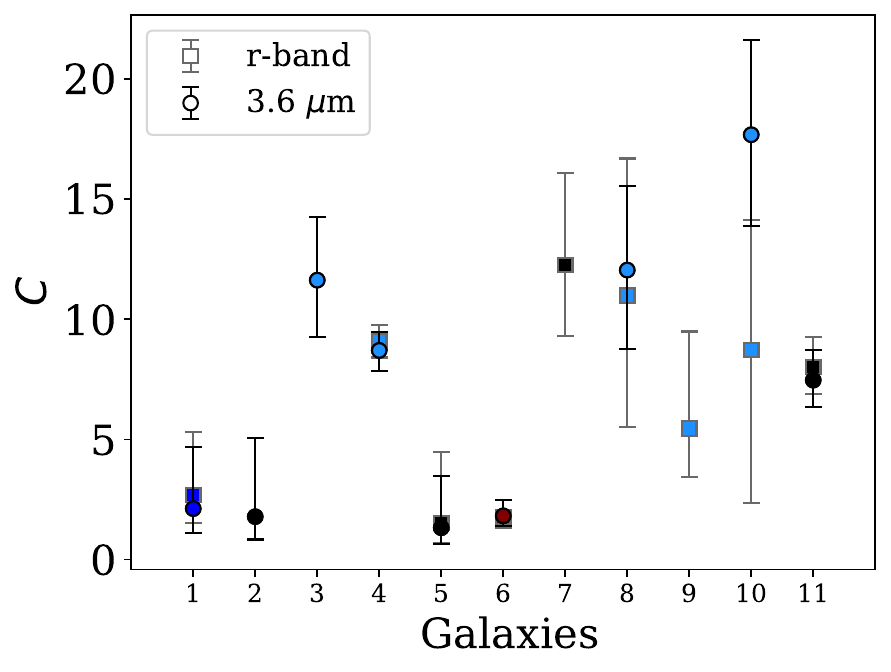}
         \caption{}
         \label{fig:mcmc_c}
     \end{subfigure}
         \begin{subfigure}[b]{0.4\textwidth}
         \centering
         \includegraphics[width=\textwidth]{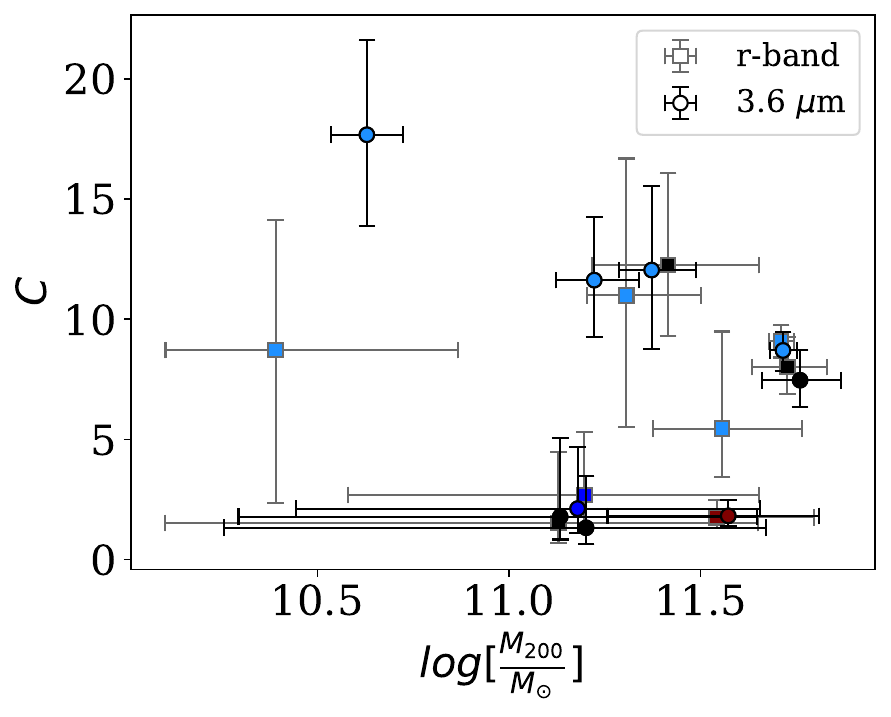}
         \caption{}
         \label{fig:mcmc_c_m200}
     \end{subfigure}
    \caption{Figure \ref{fig:mcmc_ml}, \ref{fig:mcmc_m200} and \ref{fig:mcmc_c} respectively show the distribution M/L, $M_{200}$ and $C$ of different galaxies found from the mean of the posterior distribution of the MCMC run. Figure \ref{fig:mcmc_c_m200} shows the distribution $C$ with $M_{200}$. The circles and squares denote parameters found while using 3.6 $\mu m$ and r-band data, respectively. Galaxies are colour coded according to their Hubble type mentioned in figure \ref{fig:comp_incl}.}
    \label{fig:mcm_param}
\end{figure*}

Table \ref{tab:mcmc_params} and figure \ref{fig:mcm_param} show the distribution of the free parameters found from the posterior distribution of the MCMC run of different galaxies. From figure \ref{fig:mcmc_ml}, we see that M/L for 3.6 $\mu$m are better constrained than that for r-band data as the prior for M/L for 3.6 $\mu$m is better constrained. From figure \ref{fig:mcmc_m200} and \ref{fig:mcmc_c}, we see that, despite of using different photometry, we get consistent results for the halo parameters. Figure \ref{fig:mcmc_c_m200}, i.e., the M$_{200}$-$C$ relation does not show any evident correlation. \citet{sparc_mass_model}, in their studies of mass modelling, also did not find any pattern in M$_{200}$-$C$ relation instead of using priors depending on M$_{star}$-M$_{200}$ and M$_{200}$-$C$ relation, instead they found marginal evidence of anti-correlation in this relation while using flat priors.

The total H~{\sc i} gas mass and star mass derived from different photometries are shown in table \ref{tab:mcmc_mass}. The H~{\sc i} mass is computed from H~{\sc i} line-flux following \citet{biswas2022}. For the computation of the total star mass, for each galaxy, for each photometry,  we took the total light (L) computed by adding the area under each Gaussian from MGE fitting and multiplying it with mass-to-light ratio derived from mass modelling.

\begin{table}
    \centering
    \begin{tabular}{ccc}
        \hline
        Source & 3.6-micron & r-band \\
        \hline
            NGC0784 & 0.37 & 0.58 \\
		NGC1156 & 0.19 & - \\
		NGC3027 & 0.06 & - \\
		NGC3359 & 1.50 & 1.48 \\
		NGC4068 & 0.09 & 0.05 \\
		NGC4861 & 2.73 & 2.26 \\
		NGC7292 & - & 0.88 \\
		NGC7497 & 0.63 & 1.00 \\
		NGC7610 & - & 5.25 \\
		NGC7741 & 6.87 & 4.51 \\
		NGC7800 & 1.49 & 1.12 \\
        \hline
    \end{tabular}
    \caption{List of reduced $\chi^2$ in the MCMC fitting}
    \label{tab:chisq}
\end{table}

\begin{table*}
    \centering
    \begingroup
    \renewcommand{\arraystretch}{1.5}
    \begin{tabular}{ccccccc}
    \hline
    Name & M/L(3.6 $\mu$m) & M$_{200}$ (M$_{\sun}$) (3.6 $\mu$m) & C (3.6 $\mu$m) & M/L(r-band) & M$_{200}$ (M$_{\sun}$) (r-band) & C (r-band) \\
    \hline
NGC0784 & $ 0.62 \pm^{0.18}_{0.18} $ & $ 1.5 \pm^{3.0}_{1.2} \times 10^{11} $ & $ 2.1 \pm^{2.6}_{1.0} $ & $ 0.8 \pm^{0.6}_{0.5} $ & $ 1.6 \pm^{2.9}_{1.2} \times 10^{11} $ & $ 2.7 \pm^{2.6}_{1.2} $ \\
NGC1156 & $ 0.25 \pm^{0.07}_{0.07} $ & $ 1.4 \pm^{3.1}_{1.2} \times 10^{11} $ & $ 1.8 \pm^{3.3}_{0.9} $ & - - - - & - - - - - - & - - - - \\
NGC3027 & $ 0.65 \pm^{0.19}_{0.19} $ & $ 1.7 \pm^{0.5}_{0.35} \times 10^{11} $ & $ 11.6 \pm^{2.6}_{2.4} $ & - - - - & - - - - - - & - - - - \\
NGC3359 & $ 0.04 \pm^{0.04}_{0.024} $ & $ 5.2 \pm^{0.5}_{0.4} \times 10^{11} $ & $ 8.7 \pm^{0.8}_{0.9} $ & $ 0.029 \pm^{0.028}_{0.014} $ & $ 5.1 \pm^{0.4}_{0.4} \times 10^{11} $ & $ 9.1 \pm^{0.7}_{0.7} $ \\
NGC4068 & $ 0.51 \pm^{0.16}_{0.16} $ & $ 1.6 \pm^{3.1}_{1.4} \times 10^{11} $ & $ 1.3 \pm^{2.1}_{0.7} $ & $ 0.6 \pm^{0.4}_{0.4} $ & $ 1.3 \pm^{3.1}_{1.2} \times 10^{11} $ & $ 1.5 \pm^{2.9}_{0.8} $ \\
NGC4861 & $ 0.09 \pm^{0.09}_{0.06} $ & $ 3.7 \pm^{2.7}_{1.9} \times 10^{11} $ & $ 1.8 \pm^{0.7}_{0.4} $ & $ 0.17 \pm^{0.23}_{0.12} $ & $ 3.5 \pm^{2.8}_{1.9} \times 10^{11} $ & $ 1.8 \pm^{0.7}_{0.5} $ \\
NGC7292 & - - - - & - - - - - - & - - - - & $ 2.2 \pm^{0.9}_{0.9} $ & $ 2.6 \pm^{1.9}_{1.0} \times 10^{11} $ & $ 12 \pm^{4}_{3.0} $ \\
NGC7497 & $ 0.58 \pm^{0.13}_{0.13} $ & $ 2.4 \pm^{0.7}_{0.4} \times 10^{11} $ & $ 12.0 \pm^{3.5}_{3.3} $ & $ 2.9 \pm^{0.9}_{1.0} $ & $ 2.0 \pm^{1.1}_{0.4} \times 10^{11} $ & $ 11 \pm^{6}_{5} $ \\
NGC7610 & - - - - & - - - - - - & - - - - & $ 3.1 \pm^{0.4}_{0.6} $ & $ 3.6 \pm^{2.2}_{1.2} \times 10^{11} $ & $ 5 \pm^{4}_{2.0} $ \\
NGC7741 & $ 0.99 \pm^{0.14}_{0.14} $ & $ 4.3 \pm^{1.0}_{0.8} \times 10^{10} $ & $ 18 \pm^{4}_{4} $ & $ 3.6 \pm^{0.4}_{0.4} $ & $ 2 \pm^{5}_{1.2} \times 10^{10} $ & $ 9 \pm^{5}_{6} $ \\
NGC7800 & $ 0.23 \pm^{0.15}_{0.13} $ & $ 5.7 \pm^{1.6}_{1.2} \times 10^{11} $ & $ 7.5 \pm^{1.3}_{1.1} $ & $ 0.12 \pm^{0.15}_{0.08} $ & $ 5.3 \pm^{1.4}_{1.0} \times 10^{11} $ & $ 8.0 \pm^{1.3}_{1.1} $ \\
\hline     
    \end{tabular}
    \endgroup
    \caption{Parameters found from the posterior distribution of the MCMC modelling. The first column presents the source name; the second, third and fourth column represents the mass-to-light ratio, M$_{200}$ and concentration parameter found in MCMC fitting using 3.6 $\mu m$ data, respectively; the fifth, sixth and seventh column respectively represents the mass-to-light ration, M$_{200}$ and concentration parameter found in MCMC fitting using r-band data, respectively. }
    \label{tab:mcmc_params}
\end{table*}

\begin{table*}
    \centering
    \begin{tabular}{cccc}
         \hline 
         Source & M$_{HI}$ (M$_{\sun}$) &  M$_{star}$(M$_{\sun}$) (3.6 $\mu$m) &  M$_{star}$(M$_{\sun}$) (r-band) \\
         \hline
		NGC0784 & $\left(4.22 \pm 0.06\right) \times 10^{8}$ & $\left(8.0 \pm 2.3\right) \times 10^{8}$ & $\left(2.9 \pm 2.0\right) \times 10^{8}$ \\
		NGC1156 & $\left(5.38 \pm 0.07\right) \times 10^{8}$ & $\left(1.11 \pm 0.31\right) \times 10^{9}$ & $- $ \\
		NGC3027 & $\left(5.14 \pm 0.06\right) \times 10^{9}$ & $\left(4.6 \pm 1.4\right) \times 10^{9}$ & $- $ \\
		NGC3359 & $\left(9.95 \pm 0.07\right) \times 10^{9}$ & $\left(1.9 \pm 1.8\right) \times 10^{9}$ & $\left(3.0 \pm 2.9\right) \times 10^{8}$ \\
		NGC4068 & $\left(1.284 \pm 0.035\right) \times 10^{8}$ & $\left(1.2 \pm 0.4\right) \times 10^{8}$ & $\left(7 \pm 5\right) \times 10^{7}$ \\
		NGC4861 & $\left(9.12 \pm 0.22\right) \times 10^{8}$ & $\left(6 \pm 6\right) \times 10^{7}$ & $\left(1.0 \pm 1.4\right) \times 10^{8}$ \\
		NGC7292 & $\left(5.49 \pm 0.15\right) \times 10^{8}$ & $- $ & $\left(1.2 \pm 0.5\right) \times 10^{9}$ \\
		NGC7497 & $\left(5.05 \pm 0.06\right) \times 10^{9}$ & $\left(1.32 \pm 0.29\right) \times 10^{10}$ & $\left(1.4 \pm 0.4\right) \times 10^{10}$ \\
		NGC7610 & $\left(2.07 \pm 0.05\right) \times 10^{10}$ & $- $ & $\left(1.90 \pm 0.26\right) \times 10^{10}$ \\
		NGC7741 & $\left(1.536 \pm 0.031\right) \times 10^{9}$ & $\left(6.2 \pm 0.9\right) \times 10^{9}$ & $\left(1.34 \pm 0.15\right) \times 10^{10}$ \\
		NGC7800 & $\left(3.69 \pm 0.08\right) \times 10^{9}$ & $\left(1.3 \pm 0.8\right) \times 10^{9}$ & $\left(3 \pm 4\right) \times 10^{8}$ \\
         \hline
    \end{tabular}
    \caption{H~{\sc i} mass and the star mass of the galaxies derived from different photometries.}
    \label{tab:mcmc_mass}
\end{table*}

\section{Notes on individual galaxies}
\label{sec:gal_note}
\begin{itemize}

 \item {{\bf{NGC0784}}: Mass modelling of this galaxy shows that there is little difference between the stellar velocity profile and hence the dark matter halo profile found using the r-band and 3.6 $\mu m$ data. This difference may happen due to the fact that light observed in the r-band is absorbed by the dust while observation at 3.6 $\mu m$ is not affected by it. Due to the presence of the dust, the stellar surface density profile may differ when observed through the r-band and 3.6 $\mu m$. The difference in surface density profile will result in a difference in the stellar velocity component. The detailed verification of this point is beyond the scope of this study.  } 

 \item {{\bf{NGC1156}}: For this galaxy,  in the inner radius, the stellar component is dominant, and as we go to the outer radius, the stellar component becomes weaker than the gas component, and dark matter dominates in the outermost radius. \citet{sparc_mass_model} also studied the mass model of this galaxy, but their rotation curve does not reach the flat part ( $\sim 4$ kpc), and the quality of the rotation curve is bad (Q $=3$, see \citet{sparc_rotc}), while our high-quality rotation curve goes $\sim 7$ kpc and clearly reaches the flat part.  \citet{sparc_mass_model} used different halo profiles for doing the mass modelling, but these profiles do not give unique mass modelling of this galaxy, and the variation of different components is significant. In the inner part, our result is similar to \citet{sparc_mass_model} for two cases of halo profiles and priors,  DC14 \citep{DC14_halo_profile} and Lucky13 \citep{sparc_mass_model} with priors from $\Lambda$CDM model.}

 \item {{\bf{NGC3027}}: This galaxy is primarily a dark matter dominant one. The contribution of the stellar component is higher than the gas component in the inner part, while it gets weaker in the outer part and the gas component dominates.}

 \item {{\bf{NGC3359}}: This one is also primarily a dark matter dominant galaxy, with dark matter dominating over other components across all the radius. The stellar component is comparable with the gas component in the innermost radius. However, as we go towards the outer radius, the stellar component decreases drastically, and the gas component dominates over it. This lower contribution of the star velocity component is a result of very low M/L found from the MCMC modelling for both 3.6 $\mu$m and r-band data.}

 \item {{\bf{NGC4068}}: The mass models of this galaxy using 3.6 $\mu m$ and r-band match well. The stellar component dominates in the inner radius, and in the outer radius, the dark matter dominates. \cite{sparc_mass_model} also studied the mass model of this galaxy, but their rotation curve does not reach the flat part ($\sim$ 2.5 kpc), while the rotation curve from the present analysis clearly reaches the flat part and extends $\sim$ 4.5 kpc. Besides that, the velocity at which the rotation curve reaches the flat part is lesser than the maximum rising velocity from \citet{sparc_mass_model}. However, in the inner part,  the mass models look similar to \citet{sparc_mass_model} when pseudo-Isothermal and Burkert profile is used for their halo model.}

\item {{\bf{NGC4861:}} Mass models of this galaxy show the stellar velocity profile is similar while using 3.6 $\mu m$ data and r-band data.  For both of the cases, in the inner radius, stellar and dark matter halo components dominate over the gas component; but towards the outer radius, the gas component gradually dominates over the stellar component and becomes comparable with the halo component.  }

 \item {{\bf{NGC7292}}: For this source, the stellar component is dominant in the inner radius, but as we go towards the outer radius, it decreases and becomes comparable to the gas component. In the outer radius, the dark matter dominates.}

 \item {{\bf{NGC7497}}:  Although the mass-modelling for this galaxy is similar while using 3.6 $\mu$m and r-band data, the fitting is better in the case of the 3.6 $\mu$m data. The distribution of its different components shows a similar trend as NGC1156 and NGC4068, i.e., stellar components dominates over gas and halo mass in the inner radius, then it decreases along the radius, the gas component increases along the radius and at the outer radius, dark matter dominates.  }

 \item {{\bf{NGC7610}}: Similar to NGC1156, NGC4068 and NGC7292, the stellar component dominates over the dark matter in the inner radius, and in the outer radius opposite happens.}

 \item {{\bf{NGC7741}}: There is a significant difference in stellar and hence the halo component for this galaxy while using the 3.6 $\mu$m and r-band data. The fitting of this galaxy is not good in comparison to other sources of this study. For the 3.6 $\mu$m data, it shows similar distribution as mentioned in case of NGC1156, NGC4068, NGC7292 and NGC7610, but for the r-band data,  the stellar component dominates the dark matter and gas components in all the radii.  This one is a barred, high-surface brightness galaxy with many HII regions in the bar and spiral arms. \citet{ghasp_mass_model_wise} also studied the mass models of this galaxy. They combined H$\alpha$ rotation curves from GHASP survey \citep{ghasp_rotc_epinat2008} with H~{\sc i} rotation curves from archival data. They used two dark matter profiles, i.e., pseudo-Isothermal and NFW; and the maximal disk model for the mass modelling. Our results are similar to them in two cases. First, when using only the H~{\sc i} rotation curve with the Isothermal halo and best fitting method is used to fit the data. Secondly, when using the hybrid rotation curve (combination of H$\alpha$ and H~{\sc i} rotation curves ) with Isothermal halo using maximum disk model.}

 \item {{\bf{NGC7800}}:  This galaxy is a dark matter-dominated galaxy and probably a gas-rich galaxy. The distribution of different components is similar to that of NGC3359.}
\end{itemize}

\section{Discussion}
\label{sec:discussion}
  This paper presents the details of the 3D kinematic modelling and mass modelling of a pilot sample of eleven galaxies from the GARCIA survey. To check the consistency of the derived quantities from our analysis and to validate our method of kinematic and mass modelling, we compare our results with some of the existing scaling relations.

  In this regard, we studied the following important scaling relations. Figure \ref{fig:scale_rel} shows the distribution of our sources in the M$_{gas}$-M$_{star}$, M$_{star}$-M$_{200}$ and M$_{gas}$-M$_{200}$ relations. The M$_{gas}$, shown in this figure, are derived by multiplying $1.33$ to H~{\sc i} mass to consider the contribution of Helium \citep{BFTR1}. The corresponding parameters of SPARC \citep{sparcmain2015}  galaxies are shown underneath as a consistency check of the parameters derived from our method. These parameters of SPARC galaxies are taken from \citet{sparc2016, BFTR1, sparc2019, sparc_mass_model} and the morphologies are derived from the \href{http://leda.univ-lyon1.fr/}{HyperLeda} database \citep{leda}. Additionally, we demonstrate the existing scaling relations from the literature for validation of our derived quantities: M$_{gas}$-M$_{star}$ from \citet{Parkash2018}; M$_{star}$-M$_{200}$ from  \citet{moster2013} and M$_{gas}$-M$_{200}$ from \citet{Padmanabhan2017}.  As for most of the galaxies, the mass modelling and the values of M$_{200}$ and $C$ are similar (see figure \ref{fig:mcm_param}) while using data from both the bands, and as M/L is better constrained using the 3.6 $\mu$m data, for showing these relations, we have used parameters derived while using 3.6 $\mu$m data only. For two galaxies (NGC7292 and NGC7610) for which the  3.6 $\mu$m data was not available, the parameters found using r-band data have been used. From these figures, it is clear that the quantities derived in our analysis show the same trend as seen in previous work.  In addition to that, the positions of GARCIA sources in these relations throughout different Hubble types and considering the presence/absence of bars are also in good agreement with the Hubble type and the existence of bars of the SPARC galaxies.  The trend that masses (M$_{star}$, M$_{gas}$ and M$_{200}$) of the galaxies increase following the Hubble sequence from Ir to Sd, Sc, Sb, Sa and towards elliptical, is seen in these figures. However, we do not see any correlations in these relations depending upon the presence or absence of the bar. This is probably because these relations depend on the global properties of the galaxies,  and the influence of the bar can be limited to the central regions of the galaxy.  

  To check the consistency of the different measures of velocity found from our analysis, we further investigate  the Baryonic Tully-Fisher relation.  \citet{fat} showed that there are distinct differences between 2D and 3D kinematic modelling in terms of the inclination angles and rotation velocities. With the H~{\sc i} interferometric data of 25 galaxies from LVHIS \citep{lvhis} sample,  \citet{fat} performed 2D fitting through ROTCUR \citep{rotcur1, rotcur2} task from GIPSY \citep{gipsy} and through DISKFIT \citep{diskfit,diskfit2}; and compared the results with that obtained in 3D fitting using FAT. They showed that for galaxies with low inclination, the inclination found from 3D fitting is lower than 2D fitting, and for galaxies with higher inclination, 3D fitting gives a higher inclination than 2D fitting. They found that inclinations inferred from that 3D fitting and 2D fitting have an average difference of $\sim 15^{\degree}$ for the galaxies of intermediate inclinations ($40^{\degree} \le i \le 70^{\degree}$); and there is an average difference of $\sim$ 16 kms$^{-1}$ in rotation velocity between 2D and 3D method for galaxies in intermediate inclinations \citep[see figure 7,][]{fat}. This difference in velocity and inclination prompts the consideration of testing the consistency of BTFR using 3D kinematic  data instead of 2D kinematic  data.

 We use different measures of velocity found throughout our analysis to establish the Baryonic Tully-Fisher relation and compare the results with \cite{BFTR1} (L19).  The total Baryonic mass is found following \citet{BFTR1}: M$_{bary}$ = M$_{star}$ + 1.33*M$_{HI}$, where M$_{bary}$ is the Baryonic mass; M$_{star}$ and M$_{HI}$ are respectively the star mass and H~{\sc i} mass as mentioned in table \ref{tab:mcmc_mass}. Most of the previous studies that establish this relation are based on the velocity widths from the single-dish spectra corrected for the optical inclinations  \citep[e.g,][]{McGaugh2000} . Whereas in the recent studies of BTFR, L13 corrected the single-dish spectra with inclination found from the 2D  kinematic modelling. They used rotation curves from the H${\alpha}$ and archival interferometric H~{\sc i} observation from \citet{sparc_rotc}. However, most of the H~{\sc i} rotation curves ($\sim$ 75 \%) used by L19 are derived by doing 2D kinematic modelling \citep[][etc.]{sanders_1998, Swaters_2009}. 

  We previously showed there could be a significant difference between optical inclination angle and H{~\sc i} inclination angle (figure \ref{fig:comp_incl}). Also, that the kinematics may differ substantially in terms of both velocity and inclination from 2D to 3D methods is evident from \citet{fat}'s studies.  These  {differences in estimated inclination and velocity from 3D modelling  may modify or better constrain the existing BTFR.  Here, we show the comparison of the BTFR using interferometric global H~{\sc i} spectra corrected for the kinematic inclination and single-dish global H~{\sc i} spectra corrected for the optical inclination (see figures \ref{fig:mbary_wm50}, \ref{fig:mbary_wp20}, \ref{fig:mbary_wm50_sd} and \ref{fig:mbary_wp20_sd}). We also show this relation using V$_{flat}$ and V$_{max}$ found from the 3D kinematic modelling (Figure \ref{fig:mbary_vflat} and \ref{fig:mbary_vmax}, respectively). Galaxies for which 3.6 $\mu$m data are available, we computed M$_{bary}$ using M$_{star}$ derived using 3.6 $\mu$m data only. For two galaxies (NGC7292 and NGC7610) for which the  3.6 $\mu$m data was not available, the corresponding parameter was found using r-band data.

  Figure \ref{fig:btfr} displays the corresponding results, where the associated BTFR relations with quantities from our analysis are shown as green-dashed lines. We use only galaxies with available 3.6 $\mu$m data to derive these relations for different velocities. These best-fitted lines are obtained using  the \say{lmfit} package in Python, which incorporates nonlinear least squares optimization for parameter estimation. The slopes, intercepts and scatter in our relations are shown and compared with L19 in table \ref{tab:btfr_comp}. We can see from figure \ref{fig:btfr} and table \ref{tab:btfr_comp} that there is an indication of a shallower slope in these relations. This difference in comparison to the existing BTFR may be because of the 3D kinematically estimated inclinations and velocities. We also noticed that for BTFR with kinematically modelled velocities, i.e., with V$_{flat}$ and V$_{max}$, the irregular galaxies are those that deviate the most from the existing BTFR and spirals with Hubble-type Sc, match well with the existing BTFR. We also noticed that there is no correlation of this relation with the presence/absence of the bar. However, due to the limited number of sources, we cannot conclusively determine if the change in slope or the deviations of irregular galaxies from the existing BTFR is statistically significant. With the inclusion of galaxies from the next and successive batches of GARCIA this can be verified with adequate confidence.

\begin{figure*}
    \centering
    \begin{subfigure}[b]{0.49\textwidth}
        \centering
        \includegraphics[width=\textwidth]{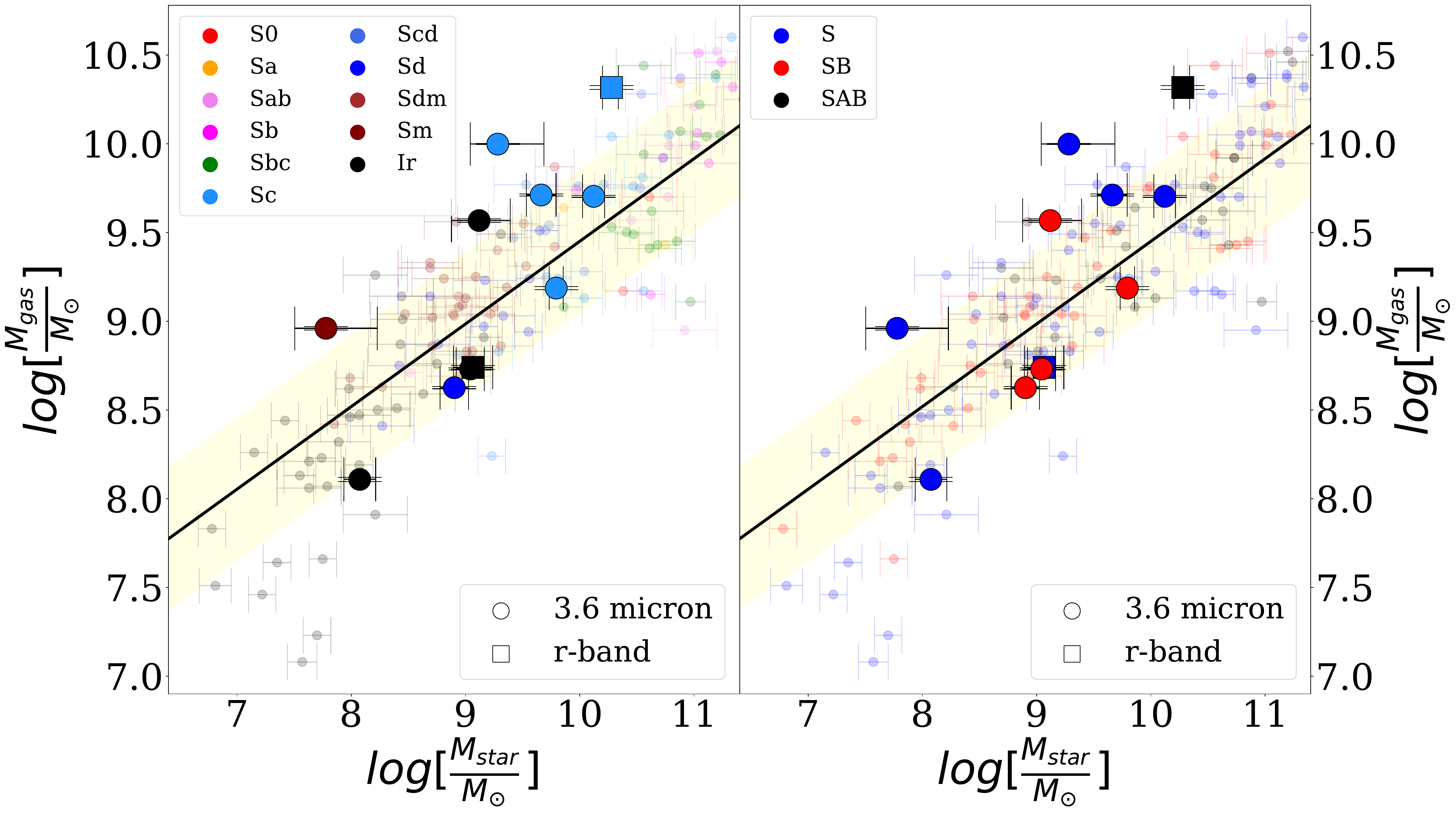}
        \caption{}
        \label{fig:mgas_mstar}
    \end{subfigure}
        \begin{subfigure}[b]{0.49\textwidth}
        \centering
        \includegraphics[width=\textwidth]{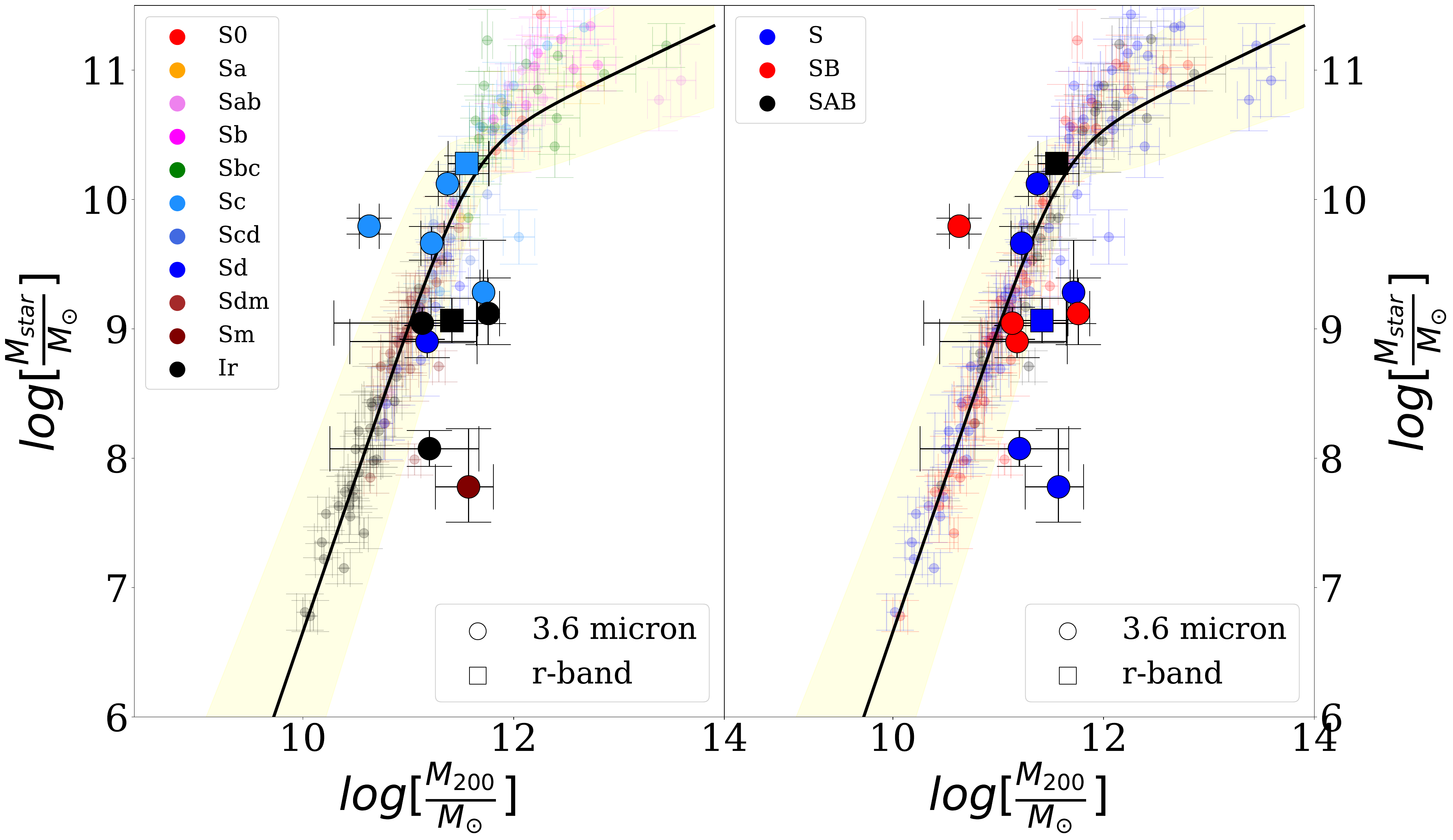}
        \caption{}
        \label{fig:mstar_m200}
    \end{subfigure}
        \begin{subfigure}[b]{0.5\textwidth}
        \centering
        \includegraphics[width=\textwidth]{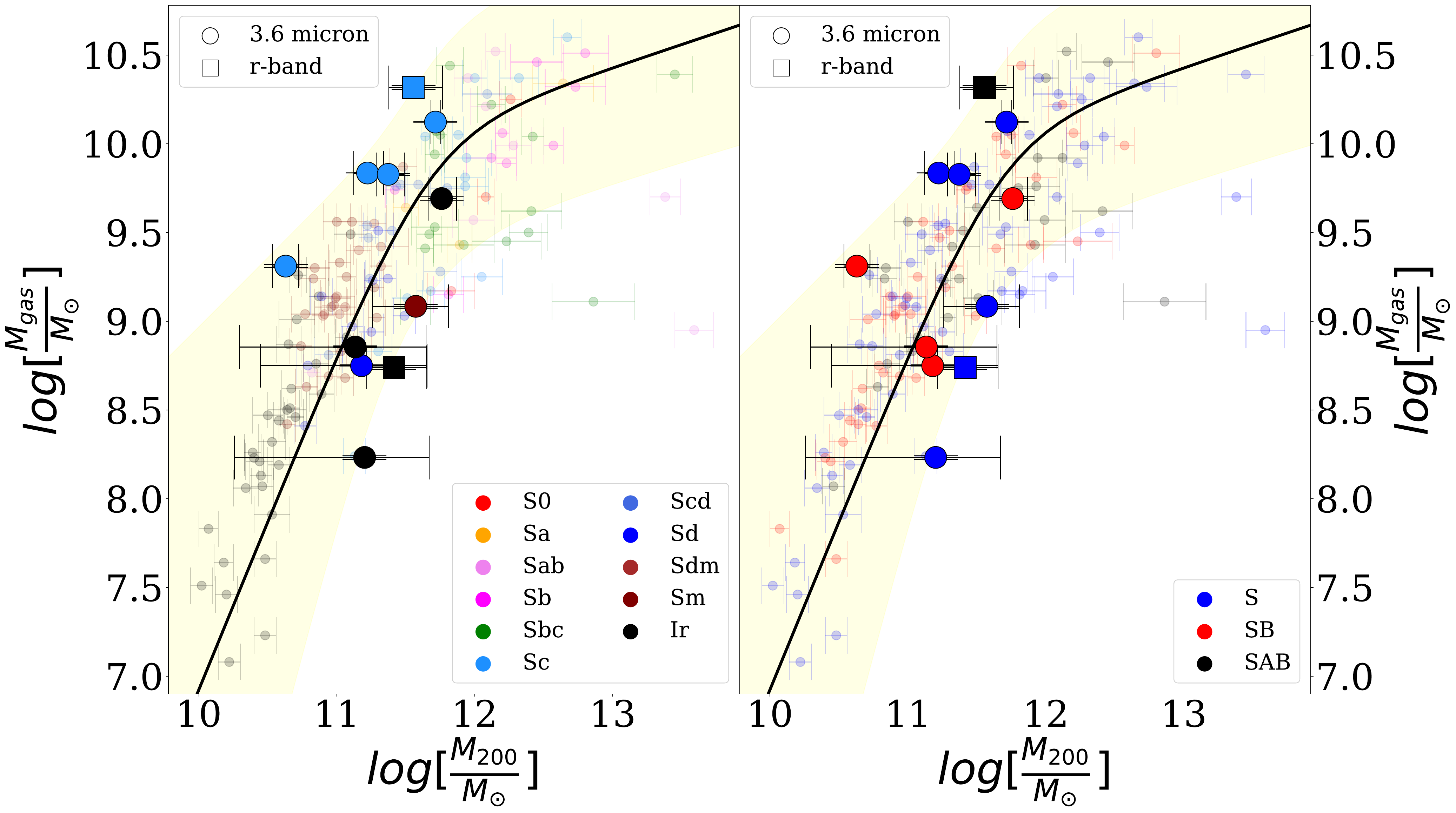}
        \caption{}
        \label{fig:mgas_m200}
    \end{subfigure}
    \caption{ Distribution of different parameters found in our analysis in  M$_{gas}$-M$_{star}$, M$_{star}$-M$_{200}$ and M$_{gas}$-M$_{200}$ relations. In the left panels of each figure, galaxies are colour-coded according to their Hubble type. Similarly, galaxies in the right-hand panel are colour-coded according to their presence of bars. The circle and square symbols represent the galaxies from GARCIA. The dots represent galaxies from the \href{http://astroweb.cwru.edu/SPARC/}{SPARC} database. The black solid line in figure \ref{fig:mgas_mstar}, \ref{fig:mstar_m200} and \ref{fig:mgas_m200} are from \citet{Parkash2018}, \citet{moster2013} and \citet{Padmanabhan2017}, respectively. The yellow shaded region in figure \ref{fig:mgas_mstar} denotes the 0.4 dex scatter around the M$_{gas}$-M$_{star}$ relation, while  in figure \ref{fig:mstar_m200} and \ref{fig:mgas_mstar}, it represents three sigma errors around the respective relations.  } 
    \label{fig:scale_rel} 
\end{figure*}

\begin{figure*}
    \centering
   \begin{subfigure}[b]{0.5\textwidth}
         \centering
         \includegraphics[width=\textwidth]{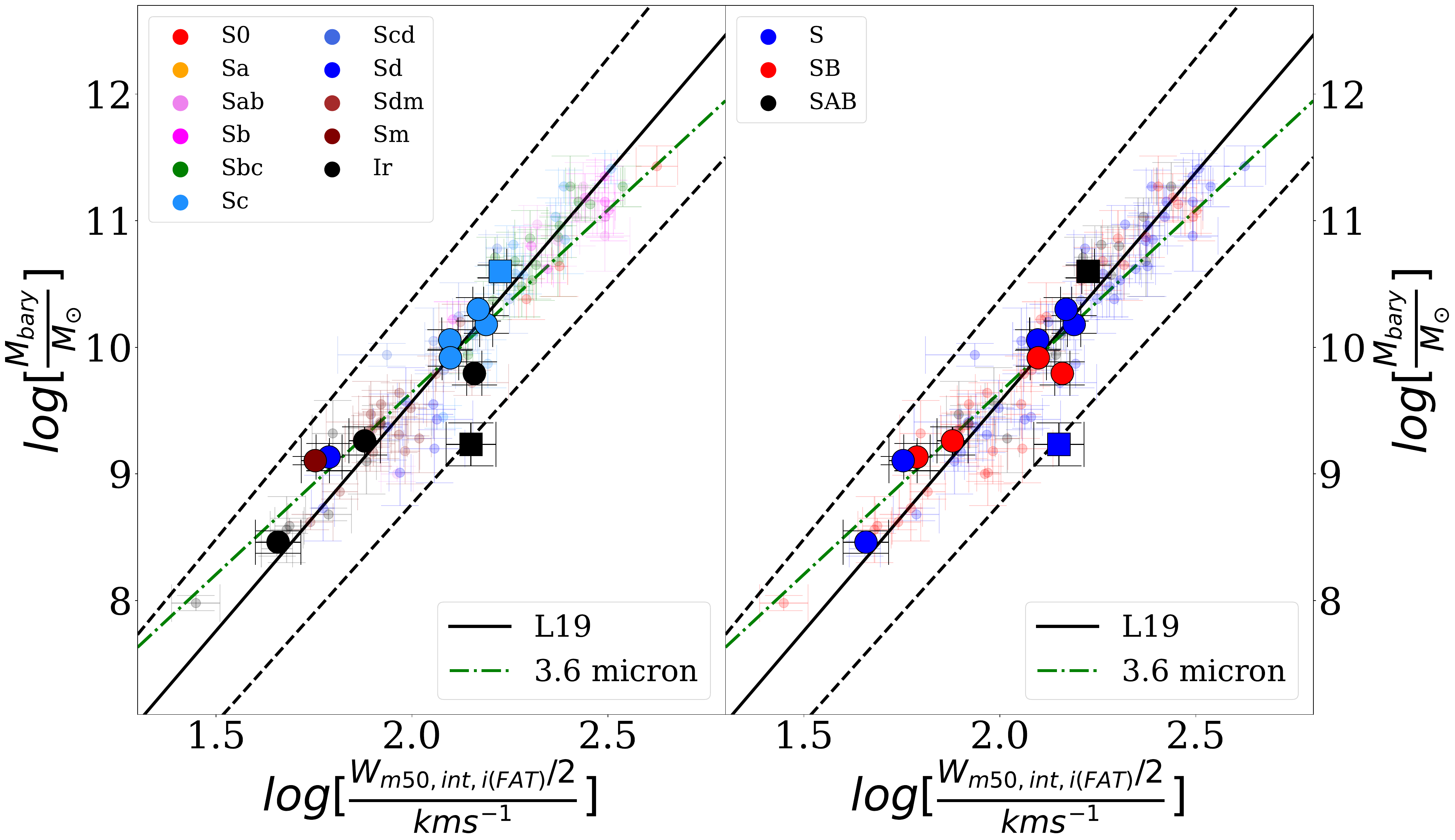}
         \caption{}
         \label{fig:mbary_wm50}
     \end{subfigure}
     \begin{subfigure}[b]{0.49\textwidth}
         \centering
         \includegraphics[width=\textwidth]{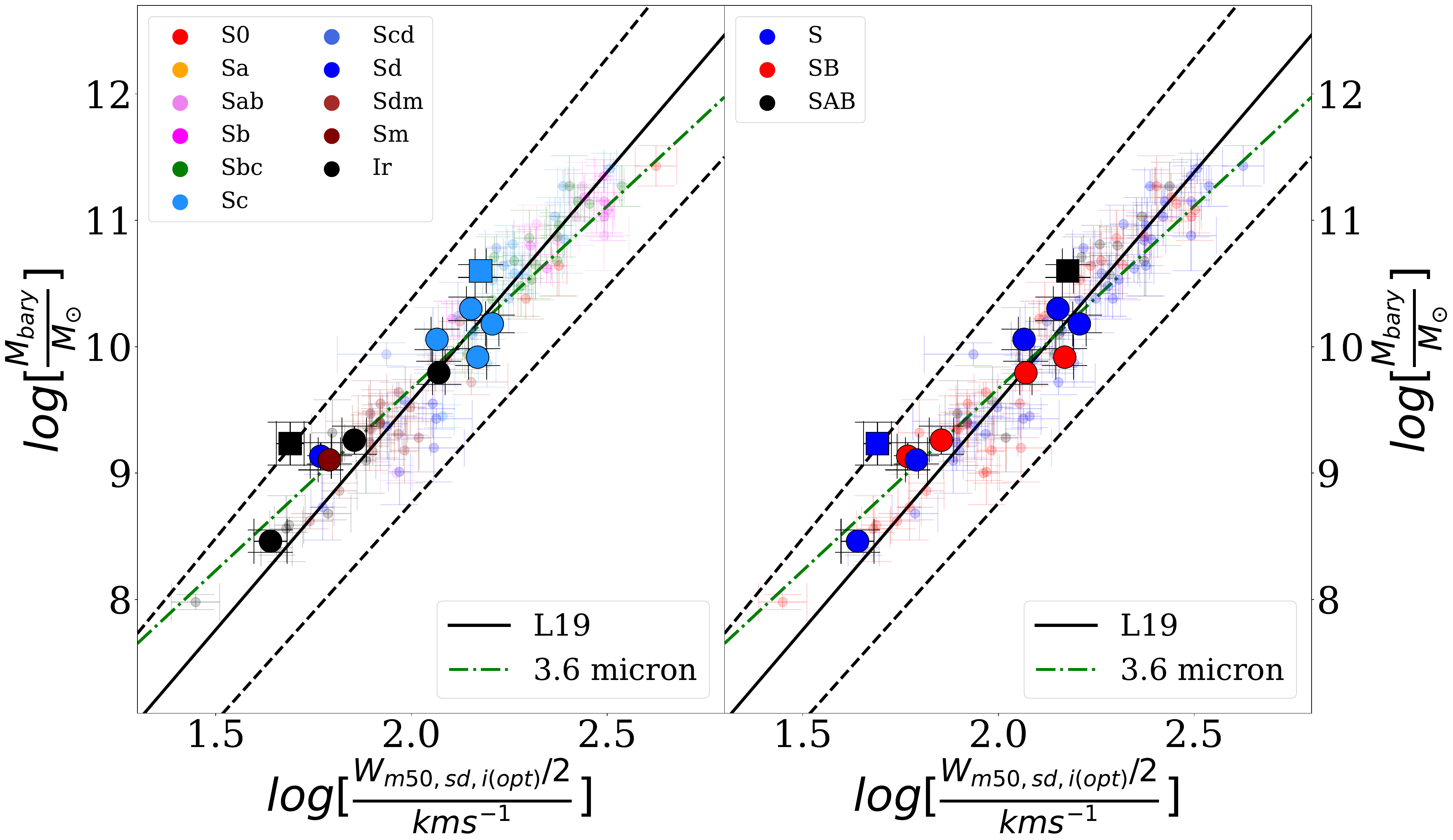}
         \caption{}
         \label{fig:mbary_wm50_sd}
     \end{subfigure}
     \begin{subfigure}[b]{0.49\textwidth}
         \centering
         \includegraphics[width=\textwidth]{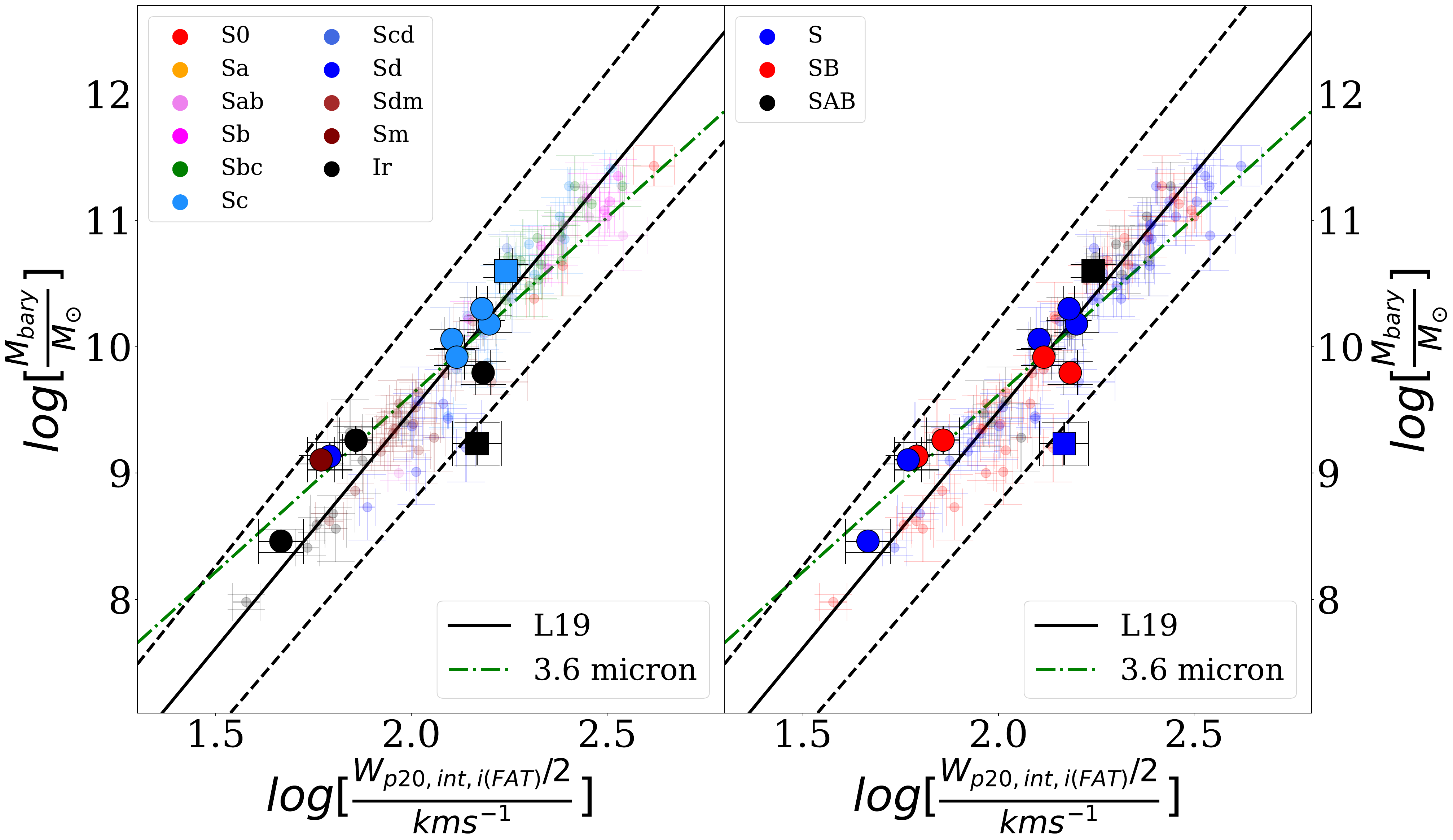}
         \caption{}
         \label{fig:mbary_wp20}
     \end{subfigure}
     \begin{subfigure}[b]{0.49\textwidth}
         \centering
         \includegraphics[width=\textwidth]{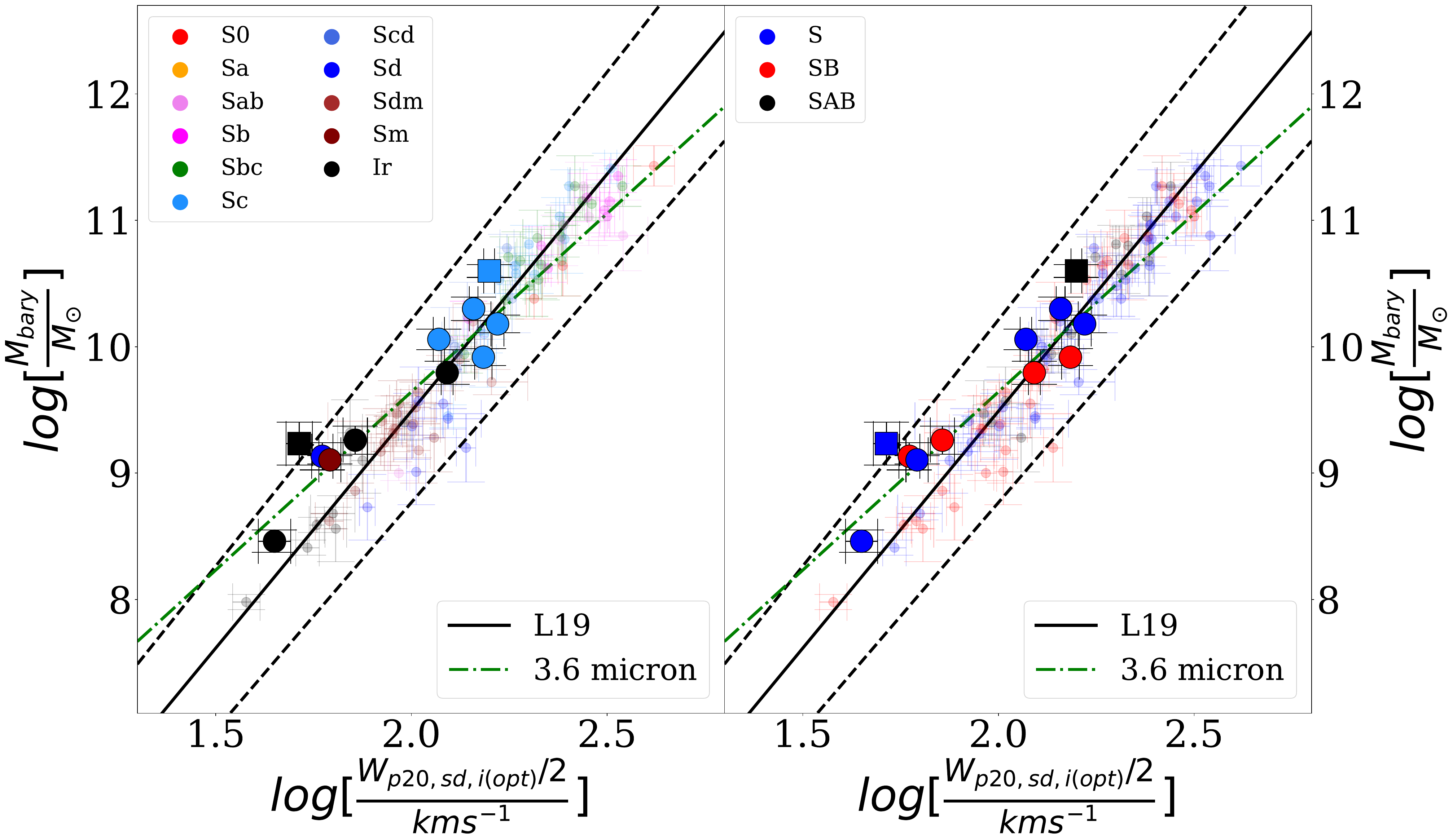}
         \caption{}
         \label{fig:mbary_wp20_sd}
     \end{subfigure}
     \begin{subfigure}[b]{0.49\textwidth}
         \centering
         \includegraphics[width=\textwidth]{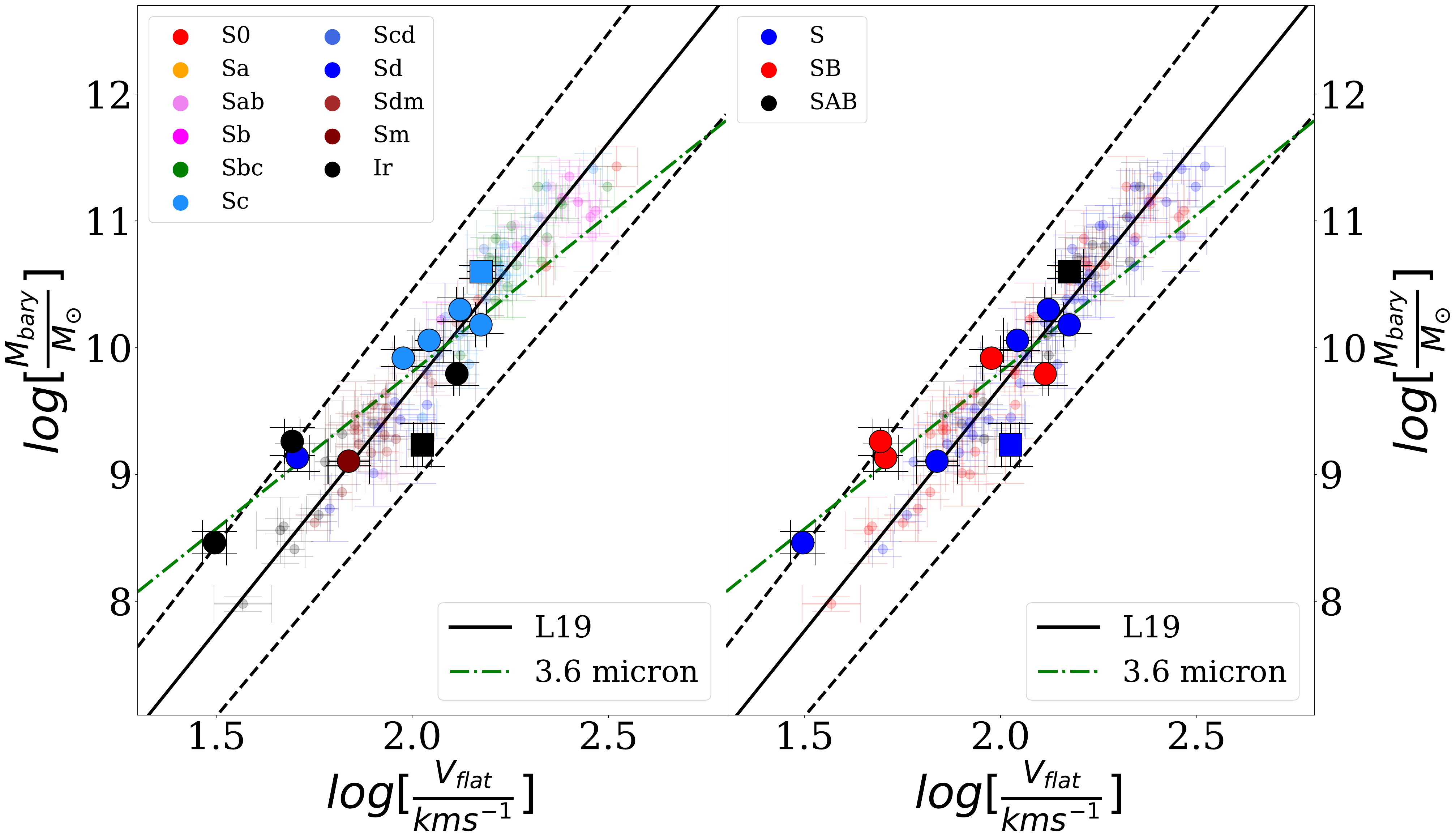}
         \caption{}
         \label{fig:mbary_vflat}
     \end{subfigure}
     \begin{subfigure}[b]{0.49\textwidth}
         \centering
         \includegraphics[width=\textwidth]{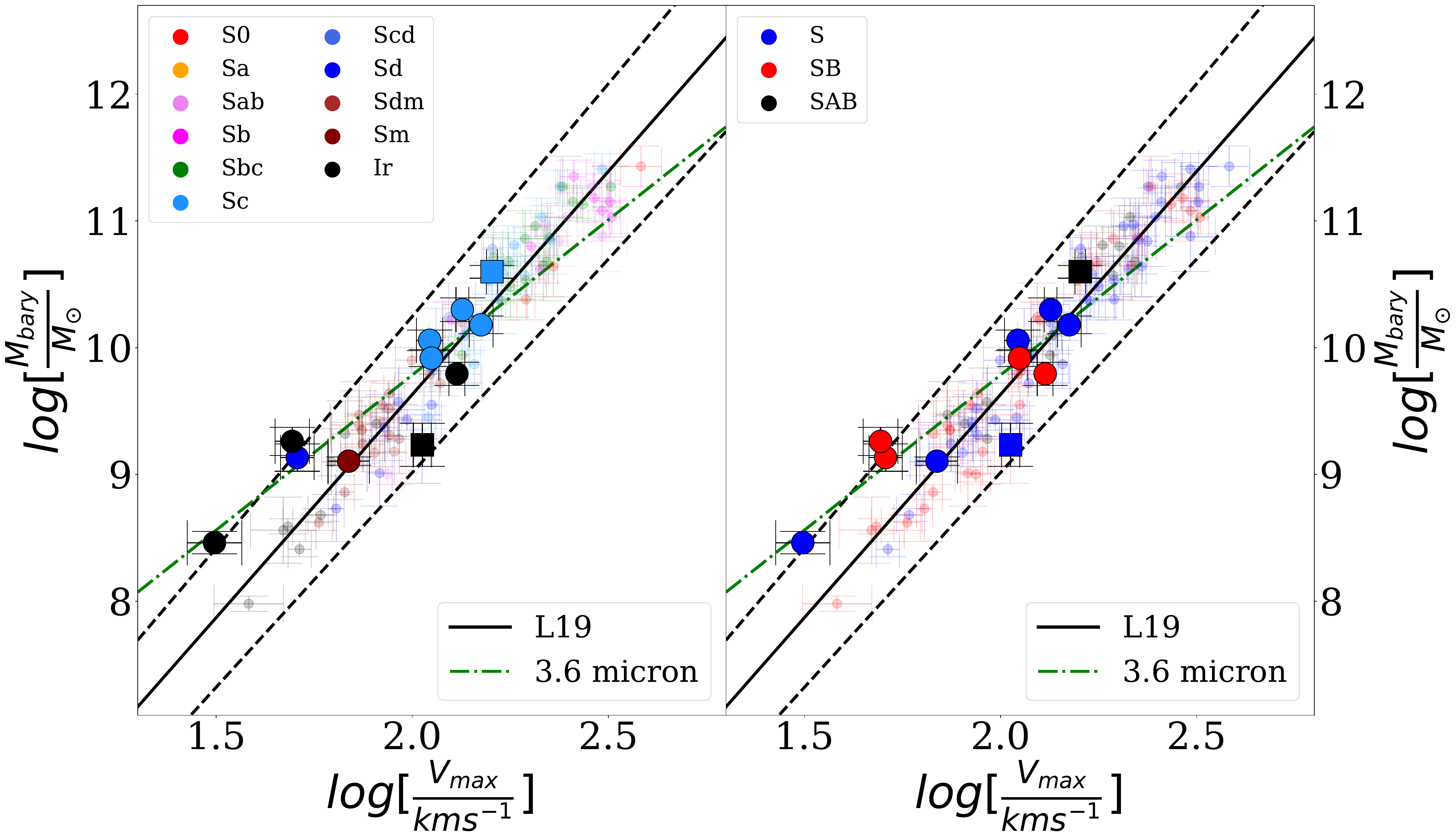}
         \caption{}
         \label{fig:mbary_vmax}
     \end{subfigure}
    \caption{  The Baryonic Tully-Fisher relation for different velocity definitions. In figure \ref{fig:mbary_wm50} and \ref{fig:mbary_wp20}, $W_{m50}$ and $W_{p20}$ are respectively from interferometric H~{\sc i} spectra corrected for the kinematic inclination angle, $i_{FAT}$; in figure \ref{fig:mbary_wm50_sd} and \ref{fig:mbary_wp20_sd}, $W_{m50}$ and $W_{p20}$ are respectively from single-dish H~{\sc i} spectra corrected for the optical inclination, $i_{opt}$. In figure \ref{fig:mbary_vflat} and \ref{fig:mbary_vmax}, $V_{flat}$ and $V_{max}$ are respectively found from the rotation curve obtained using FAT. The solid black line represents the best-fitted line from L19. The black dotted lines represent the three sigma errors in fitting around the solid line. The green dashed line represents the best-fitted line from our analysis considering only those galaxies for which the 3.6 $\mu$m data is available.  In the left panels of each figure, galaxies are colour-coded according to their Hubble type. Similarly, galaxies in the right-hand panel are colour-coded according to their presence of bars. The circle and square symbols represent the galaxies from GARCIA. The dots represent galaxies from the \href{http://astroweb.cwru.edu/SPARC/}{SPARC} database.  }
    \label{fig:btfr}
\end{figure*}

\begin{table}
    \centering
    \renewcommand{\arraystretch}{0.5}
    \setlength{\tabcolsep}{5pt}
    \caption{  Comparison of the slope, intercept and scatter from different velocities.  }
    \begin{tabular}{ccccc}
    \hline
    Y-axis & & Slope & Intercept & Scatter \\
    \hline
 		              & 3.6 $\mu$m & 2.48 $\pm$ 0.96 & 4.85 $\pm$ 1.86 & 0.06 $\pm$ 0.12 \\
            V$_{flat}$ &    &   &   &   \\     
		               & L19 & 3.85 $\pm$ 0.27 & 1.99 $\pm$ 0.54 & 0.03 $\pm$ 0.03 \\
            \hline
    		         & 3.6 $\mu$m & 2.45 $\pm$ 0.89 & 4.89 $\pm$ 1.72 & 0.06 $\pm$ 0.12 \\
            V$_{max}$ &    &   &   &   \\  
		           & L19 & 3.75 $\pm$ 0.21 & 2.59 $\pm$ 0.45 & 0.04 $\pm$ 0.03 \\
            \hline
		                       & 3.6 $\mu$m & 2.88 $\pm$ 0.93 & 3.89 $\pm$ 1.82 & 0.04 $\pm$ 0.09 \\
            W$_{m50,int,i_{FAT}}$ &    &   &   &   \\
		                       & L19 & 3.62 $\pm$ 0.27 & 2.33 $\pm$ 0.60 & 0.04 $\pm$ 0.03 \\
            \hline
                                 & 3.6 $\mu$m & 2.88 $\pm$ 0.87 & 3.91 $\pm$ 1.7 & 0.04 $\pm$ 0.09 \\
            W$_{m50,sd,i_{opt}}$ &    &   &   &   \\
		                       & L19 & 3.62 $\pm$ 0.27 & 2.33 $\pm$ 0.60 & 0.04 $\pm$ 0.03 \\
            \hline
		                       & 3.6 $\mu$m & 2.8 $\pm$ 0.93 & 4.01 $\pm$ 1.88 & 0.04 $\pm$ 0.30 \\
            W$_{p20,int,i_{FAT}}$ &    &   &   &   \\
		                       & L19 & 3.75 $\pm$ 0.24 & 1.99 $\pm$ 0.54 & 0.04 $\pm$ 0.03 \\
            \hline
		                       & 3.6 $\mu$m & 2.82 $\pm$ 0.9 & 4.0 $\pm$ 1.81 & 0.05 $\pm$ 0.09 \\
            W$_{p20,sd,i_{opt}}$ &    &   &   &   \\
		                       & L19 & 3.75 $\pm$ 0.24 & 1.99 $\pm$ 0.54 & 0.04 $\pm$ 0.03 \\        
    \hline     
    \end{tabular}
    \label{tab:btfr_comp}
\end{table}

\section{Conclusion}
\label{sec:conclusion}

Our study shows that the mass models of the galaxies are similar when using infrared 3.6 $\mu m$ data and optical r-band data for most of the cases. Significant differences in the stellar component and halo component are found  for one of the sources (NGC7741).   The parameters found from the analysis are in good agreement with the existing scaling relations. To check if the 3D-modelled data products cause any significant difference or not in the existing scaling relations, it requires a larger number of sources. Also, a bigger sample of galaxies is required to find if there are any dependencies in the morphology of galaxies in these scaling relations. With the analyzed data products of the GARCIA full sample, these questions can be answered satisfactorily. 

Besides that, as a part of the next papers of this series, we will explore other science cases as mentioned in \citet{biswas2022} e.g., finding the vertical scale height \citep{naren_hi_height}, studying the volumetric star-formation law \citep{volmetric_star_form}, dynamical modelling etc. Further, the new data products from the GARCIA sample will also be presented in our subsequent papers.  

The H~{\sc i} rotation curve for most of the sources from our analysis clearly reaches the flat part where the dark matter dominates. Although, the rising inner part, where the rotation curve changes the most, is less constrained because of the limited beam size of H~{\sc i} observations. The modelling of the rotation curve can be better constrained if we combine it with the H${\alpha}$ rotation curve or rotation curve found from the stellar dynamics as done by \citet{ghasp_mass_model_wise} and \citet{Teodoro2022_3Dmass_model_mcmc}. Unfortunately, for most of the GARCIA galaxies, the inner rotation curve is not available. Also, for this current sample of galaxies, none of the inner rotation curves is available publicly.

However, our recent GMRT observations of CALIFA (The Calar Alto Legacy Integral Field Area survey \citep{califa1_2012,sanchez2016}) galaxies as a part of the MasQue: Mass Modelling and Star-formation Quenching of Nearby Galaxies  ( PI: V. Kalinova), have the inner circular velocity curve (CVC) available from stellar dynamics from \citet{kalinova2017}. CALIFA survey is an optical integral field unit (IFU) survey that imaged a representative sample of $\sim1000$ Sloan Digital Sky Survey (SDSS) galaxies. This survey provides kpc-resolved stellar property maps and covers galaxies in  different stages of their evolution \citep{kalinova2021, kalinova2022}. Some of these galaxies have their follow-up CO line observation by Extragalactic Database for Galaxy Evolution (EDGE, \citeauthor{bolatto2017} \citeyear{bolatto2017}) collaboration using The Combined Array for Research in Millimeter-wave Astronomy (CARMA), The Atacama Pathfinder Experiment (APEX, \citeauthor{apex_colombo2020} \citeyear{apex_colombo2020}), and Atacama Compact Array (ACA). EDGE  also acquired single-dish H~{\sc i} data of all the proposed targets with the GBT and from the archives. In our next paper, we will utilize the potential of this data set and combine the H~{\sc i} rotation curve with CVC and perform the mass modelling of 10+ CALIFA galaxies considering a mass-to-light ratio constant with radius. Further, the mass-to-light ratio of the galaxies, using either infrared or optical data, may vary radially. So, if we consider the radial variation of the mass-to-light ratio, the modelling can be even better constrained; this idea will be explored in detail with the combined rotation curve in our following papers. 

\section*{Acknowledgements}
We convey our heartfelt acknowledgements to Dr Ling Zhu, Dr. Sharon van der Wel, Dr Jesus Falcon-Barroso, Dr Michele Cappellari and Dr Sushma Kurapati for helping us in different ways to derive this work. Sushma helped us to learn the way how FAT works. Ling and Sharon gave us valuable insights into understanding how to use the S4G data for MGE fitting. Jesus and Michele helped us understand the MGE fitting procedure of the Spitzer-IRAC-3.6 micron data.  We also thank the anonymous reviewers for their suggestions and comments that have helped to improve this paper.  Moreover, this work has made use of the SPITZER and SDSS databases. The Spitzer Space Telescope was operated by the Jet Propulsion Laboratory, California Institute of Technology under a contract with NASA. Support for this work was provided by an award issued by JPL/Caltech. For the Sloan Digital Sky Survey (SDSS), funding has been provided by the Alfred P. Sloan Foundation; the participating institutions are the National Aeronautics and Space Administration, the National Science Foundation, the U.S. Department of Energy, the Japanese Monbukagakusho, and the Max Planck Society. This research has also utilized the NASA/IPAC Extragalactic Database (NED), operated by the Jet Propulsion Laboratory, California Institute of Technology, under contract with the National Aeronautics and Space Administration. Additionally, we extensively utilized the archival data from the GMRT online archive in this work. GMRT is run by the National Centre for Radio Astrophysics of the Tata Institute of Fundamental Research. We truly appreciate the facility of the GMRT online archive and sincerely thank the entire GMRT team.

%%%%%%%%%%%%%%%%%%%%%%%%%%%%%%%%%%%%%%%%%%%%%%%%%%
\section*{Data Availability}

All the derived quantities and models produced in this study will be shared at reasonable request to the corresponding author.



\bibliographystyle{mnras}
\bibliography{example} % if your bibtex file is called example.bib

%\begin{thebibliography}{99}
%\end{thebibliography}

%%%%%%%%%%%%%%%%%%%%%%%%%%%%%%%%%%%%%%%%%%%%%%%%%%
%%%%%%%%%%%%%%%%% APPENDICES %%%%%%%%%%%%%%%%%%%%%
\appendix
%%%%%%%%%%%%%%%%%%%%%%%%%%%%%%%%%%%%%%%%%%%%%%%%%%
%%%%%%%%%%%%%%%%%%%%%%%%%%%%%%%%%%%%%%%%%%%%%%%%%%

\section{Finding the gas velocity from gas surface density}
\label{vgas_accr}
As mentioned in section \ref{gas_compnt}, for an axisymmetric razor-thin disk ( where the thickness of disk tends to zero) of surface density $\Sigma (R) $, the velocity can be found through equation \ref{eqn:vgas}. As the Bessel functions are highly oscillatory, the integration in equation \ref{eqn:vgas}  will give inaccurate results if we follow the usual methods (use of the Trapezoidal rule or Simpson's 1/3 rule) of doing numerical integration. We followed the procedure described below to check the accuracy of our method of computing this integration. We consider an exponential disk with $\Sigma = \Sigma(0) \exp^{-R/R_d}$, where $R_d$ is the scale radius and $\Sigma(0)$ is the surface density at $R=0$. Equation \ref{eqn:vgas} can be solved analytically for this type of surface density. The following equation gives the corresponding solution:

\begin{align}
  \label{eq:vel_comp}
  \begin{split}
    v^2(R) = \pi G \Sigma(0) & \frac{R^2}{2R_d} \times  \\ & \left[I_0\left( \frac{R}{2R_d} \right) K_0 \left( \frac{R}{2R_d} \right) -  I_1\left( \frac{R}{2R_d} \right) K_1\left( \frac{R}{2R_d} \right)  \right] ,
\end{split}
\end{align}

Where $I_m(\cdot)$ and $K_m(\cdot)$ are respectively modified Bessel functions of the first and second kind with order $m$, we first consider the disk to have a radius of 200$^{\prime \prime}$ and then discretize it into 10 points. This gives us a ring size of $\sim 20^{\prime \prime}$, which is comparable to the beam size of our analysis. We use this discretized $\Sigma$ to determine the velocity using our method. The result is compared with the analytical solution, which is discretized in the same manner. The top panel of figure \ref{fig:vgas_accr} shows the comparison between the velocities found from our method and the analytical solution, and the bottom panel shows the difference between them. This figure suggests that this method is suitable for finding the gas velocity accurately in the outer radius, and the difference in the velocity with the analytical one is small in the inner radius. The result justifies our method of finding the gas velocity from the surface density.

\begin{figure}
    \centering
    \includegraphics[height=5cm]{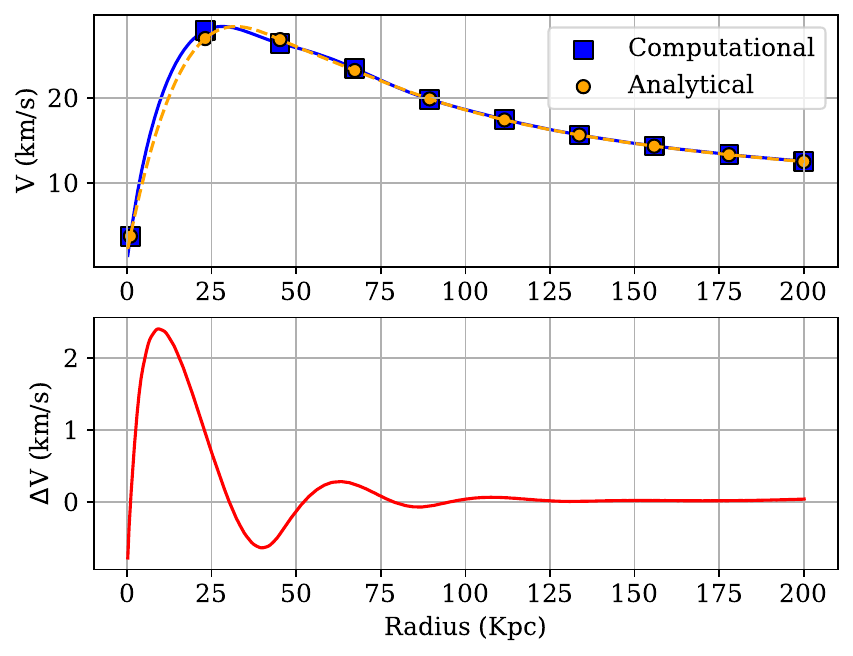}
    \caption{Accuracy of computing the gas velocity from surface density. Top panel: Yellow circular points denote the gas profile given by equation \ref{eq:vel_comp} i.e.,  by solving equation \ref{eqn:vgas} analytically using a surface density profile of exponential form, and the blue square points denote numerically obtained gas profile from the same equation, i.e., equation \ref{eqn:vgas} using the same form of surface density via our method. Down panel: The difference between the numerically and the analytically obtained gas profiles.  }
    \label{fig:vgas_accr}
\end{figure}

% Don't change these lines
\bsp	% typesetting comment
\label{lastpage}

% End of mnras_template.tex

\end{document}